\documentclass[sigconf,screen]{acmart}
\pdfoutput=1  

\usepackage[braket]{qcircuit}
\usepackage{amsmath,amssymb,amsfonts}
\usepackage{graphicx}
\usepackage{xcolor}
\usepackage{svg}
\usepackage{xparse}
\usepackage{balance}
\usepackage{fancyhdr}

\usepackage{tikz}
\usepackage{pgfplots}
\usetikzlibrary{patterns}
\usetikzlibrary{arrows}
\usetikzlibrary{external}
\usetikzlibrary{shapes}
\usetikzlibrary{fadings,shapes.arrows,shadows}
\usetikzlibrary{shadows.blur}
\usetikzlibrary{shapes.symbols}
\usetikzlibrary{shapes.multipart}
\usetikzlibrary{positioning}

\pgfplotsset{compat=1.16}

\newcommand{\hide}[1]{}
\tikzfading[name=arrowfading, top color=transparent!0, bottom color=transparent!95]
\tikzset{arrowfill/.style={#1,
    }}
\tikzset{arrowstyle1/.style n args={0}{%
    arrowfill={top color=plotBlue!20!white,bottom color=plotBlue!70!white, shape border rotate=270},
    single arrow,
    minimum height=1.4cm,
    minimum width=1.6cm,
    single arrow head extend=0.2cm,}}
\tikzset{arrowstyle2/.style n args={0}{%
    arrowfill={top color=plotGreen!20!white,bottom color=plotGreen!70!white, shape border rotate=270},
    single arrow,
    minimum height=1.4cm,
    minimum width=1.6cm,
    single arrow head extend=0.2cm,}}
\tikzset{arrowstyle3/.style n args={0}{%
    arrowfill={top color=plotRed!20!white,bottom color=plotRed!70!white, shape border rotate=270},
    single arrow,
    minimum height=1.4cm,
    minimum width=1.6cm,
    single arrow head extend=0.2cm,}}
\definecolor{plotBlue}{rgb}{0.37894736842105264, 0.6947368421052632, 0.9947368421052631}
\definecolor{plotGreen}{rgb}{0.5894736842105263, 0.8894736842105263, 0.23684210526315788}
\definecolor{plotRed}{rgb}{0.9473684210526315, 0.29473684210526313, 0.531578947368421}

\setcopyright{acmlicensed}
\acmPrice{15.00}
\acmDOI{10.1145/3445814.3446718}
\acmYear{2021}
\copyrightyear{2021}
\acmSubmissionID{asplos21main-p276-p}
\acmISBN{978-1-4503-8317-2/21/04}
\acmConference[ASPLOS '21]{Proceedings of the 26th ACM International Conference on Architectural Support for Programming Languages and Operating Systems}{April 19--23, 2021}{Virtual, USA}
\acmBooktitle{Proceedings of the 26th ACM International Conference on Architectural Support for Programming Languages and Operating Systems (ASPLOS '21), April 19--23, 2021, Virtual, USA}

\fancypagestyle{firststyle}{%
  \fancyhf{}%
  
  \fancyfoot[C]{\vspace{2em}\scriptsize%
    \copyright The authors 2021. This is the author's version of the work. It is posted here for your personal use. Not for redistribution. The definitive version was published in \textit{Proceedings of the 26th ACM International Conference on Architectural Support for Programming Languages and Operating Systems (ASPLOS ’21)}, \url{https://doi.org/10.1145/3445814.3446718}.%
  }%
}
\fancypagestyle{mynumbered}{%
  
  \fancyfoot[C]{\vspace{2em}\thepage}%
}

\begin{document}

\title[Orchestrated Trios: Compiling for Efficient Communication in Quantum Programs with 3-Qubit Gates]{Orchestrated Trios: Compiling for Efficient Communication\\in Quantum Programs with 3-Qubit Gates}

\AfterPreamble{\hypersetup{
  pdfauthor={TODO},
  pdftitle={Orchestrated Trios: Compiling for Efficient Communication in Quantum Programs with 3-Qubit Gates},
  pdfsubject={TODO},
}}

\settopmatter{authorsperrow=4}
\author{Casey Duckering}
\orcid{0000-0002-4656-9644}
\affiliation{%
  \institution{University of Chicago}
  \streetaddress{5730 S. Ellis Avenue}
  \city{Chicago}
  \state{Illinois}
  \country{USA}
  \postcode{60637}
}
\author{Jonathan M. Baker}
\orcid{0000-0002-0775-8274}
\affiliation{%
  \institution{University of Chicago}
  \streetaddress{5730 S. Ellis Avenue}
  \city{Chicago}
  \state{Illinois}
  \country{USA}
  \postcode{60637}
}
\author{Andrew Litteken}
\orcid{0000-0001-5676-1747}
\affiliation{%
  \institution{University of Chicago}
  \streetaddress{5730 S. Ellis Avenue}
  \city{Chicago}
  \state{Illinois}
  \country{USA}
  \postcode{60637}
}
\author{Frederic T. Chong}
\orcid{0000-0001-9282-4645}
\affiliation{%
  \institution{University of Chicago}
  \streetaddress{5730 S. Ellis Avenue}
  \city{Chicago}
  \state{Illinois}
  \country{USA}
  \postcode{60637}
}

\begin{abstract}

Current quantum computers are especially error prone and require high levels of optimization to reduce operation counts and maximize the probability the compiled program will succeed.  These computers only support operations decomposed into one- and two-qubit gates and only two-qubit gates between physically connected pairs of qubits.  Typical compilers first decompose operations, then route data to connected qubits.  We propose a new compiler structure, Orchestrated Trios, that first decomposes to the three-qubit Toffoli, routes the inputs of the higher-level Toffoli operations to groups of nearby qubits, then finishes decomposition to hardware-supported gates.

This significantly reduces communication overhead by giving the routing pass access to the higher-level structure of the circuit instead of discarding it.  A second benefit is the ability to now select an architecture-tuned Toffoli decomposition such as the 8-CNOT Toffoli for the specific hardware qubits now known after the routing pass.  We perform real experiments on IBM Johannesburg showing an average 35\% decrease in two-qubit gate count and 23\% increase in success rate of a single Toffoli over Qiskit.  We additionally compile many near-term benchmark algorithms showing an average 344\% increase in (or 4.44x) simulated success rate on the Johannesburg architecture and compare with other architecture types.

\end{abstract}



\maketitle
\vspace{100em}  
\pagestyle{mynumbered}
\thispagestyle{firststyle}

\section{Introduction}

\begin{figure}
    \centering
    \begin{minipage}[][][b]{\columnwidth}
        \centering
        \scalebox{1.5}{\begin{tikzpicture} [node distance=1.2cm,scale=0.6, every node/.style={scale=0.6}]
    \node[fill={rgb,190:red,72;green,132;blue,189},draw,circle,line width=0.5pt,minimum size=0.75cm,] (0) {0};
    \node[fill={rgb,190:red,72;green,132;blue,189},draw,circle,line width=0.5pt,minimum size=0.75cm,right of=0] (1) {1};
    \node[fill={rgb,190:red,180;green,56;blue,101},draw,circle,line width=0.5pt,minimum size=0.75cm,right of=1] (2) {2};
    \node[fill={rgb,190:red,72;green,132;blue,189},draw,circle,line width=0.5pt,minimum size=0.75cm,right of=2] (3) {3};
    \node[fill={rgb,190:red,72;green,132;blue,189},draw,circle,line width=0.5pt,minimum size=0.75cm,right of=3] (4) {4};
    \node[fill={rgb,190:red,72;green,132;blue,189},draw,circle,line width=0.5pt,minimum size=0.75cm,below of=0] (5) {5};
    \node[fill={rgb,190:red,180;green,56;blue,101},draw,circle,line width=0.5pt,minimum size=0.75cm,below of=1] (6) {6};
    \node[fill={rgb,190:red,72;green,132;blue,189},draw,circle,line width=0.5pt,minimum size=0.75cm,below of=2] (7) {7};
    \node[fill={rgb,190:red,72;green,132;blue,189},draw,circle,line width=0.5pt,minimum size=0.75cm,below of=3] (8) {8};
    \node[fill={rgb,190:red,72;green,132;blue,189},draw,circle,line width=0.5pt,minimum size=0.75cm,below of=4] (9) {9};
    \node[fill={rgb,190:red,72;green,132;blue,189},draw,circle,line width=0.5pt,minimum size=0.75cm,below of=5] (10) {10};
    \node[fill={rgb,190:red,72;green,132;blue,189},draw,circle,line width=0.5pt,minimum size=0.75cm,below of=6] (11) {11};
    \node[fill={rgb,190:red,72;green,132;blue,189},draw,circle,line width=0.5pt,minimum size=0.75cm,below of=7] (12) {12};
    \node[fill={rgb,190:red,72;green,132;blue,189},draw,circle,line width=0.5pt,minimum size=0.75cm,below of=8] (13) {13};
    \node[fill={rgb,190:red,72;green,132;blue,189},draw,circle,line width=0.5pt,minimum size=0.75cm,below of=9] (14) {14};
    \node[fill={rgb,190:red,72;green,132;blue,189},draw,circle,line width=0.5pt,minimum size=0.75cm,below of=10] (15) {15};
    \node[fill={rgb,190:red,72;green,132;blue,189},draw,circle,line width=0.5pt,minimum size=0.75cm,below of=11] (16) {16};
    \node[fill={rgb,190:red,72;green,132;blue,189},draw,circle,line width=0.5pt,minimum size=0.75cm,below of=12] (17) {17};
    \node[fill={rgb,190:red,72;green,132;blue,189},draw,circle,line width=0.5pt,minimum size=0.75cm,below of=13] (18) {18};
    \node[fill={rgb,190:red,180;green,56;blue,101},draw,circle,line width=0.5pt,minimum size=0.75cm,below of=14] (19) {19};

    \draw
        (0) edge[line width=2pt] node{} (1)
        (1) edge[line width=2pt] node{} (2)
        (2) edge[line width=2pt] node{} (3)
        (3) edge[line width=2pt] node{} (4)
        (0) edge[line width=2pt] node{} (5)
        (4) edge[line width=2pt] node{} (9)
        (5) edge[line width=2pt] node{} (6)
        (6) edge[line width=2pt] node{} (7)
        (7) edge[line width=2pt] node{} (8)
        (8) edge[line width=2pt] node{} (9)
        (5) edge[line width=2pt] node{} (10)
        (9) edge[line width=2pt] node{} (14)
        (7) edge[line width=2pt] node{} (12)
        (10) edge[line width=2pt] node{} (11)
        (11) edge[line width=2pt] node{} (12)
        (12) edge[line width=2pt] node{} (13)
        (13) edge[line width=2pt] node{} (14)
        (10) edge[line width=2pt] node{} (15)
        (14) edge[line width=2pt] node{} (19)
        (15) edge[line width=2pt] node{} (16)
        (16) edge[line width=2pt] node{} (17)
        (17) edge[line width=2pt] node{} (18)
        (18) edge[line width=2pt] node{} (19)
        
        (19) edge[line width=1pt,->,bend left=50,below] node{1} (18)
        (6) edge[line width=1pt,->,bend right=50,above] node{1} (5)
        (2) edge[line width=1pt,->,bend right=50,above] node{1} (1)
        
        (18) edge[line width=1pt,->,bend left=50,below] node{2} (17)
        (5) edge[line width=1pt,->,bend left=30,left] node{2} (0)
        (17) edge[line width=1pt,->,bend left=50,below] node{3} (16)
        
        (0) edge[line width=1pt,->,bend right=80,left] node{4, 10} (5)
        
        (5) edge[line width=1pt,->,bend right=80,left] node{5, 9} (10)
        (1) edge[line width=1pt,->,bend right=50,above] node{5} (0)
        
        (10) edge[line width=1pt,->,bend right=80,left] node{6, 10} (15)
        (15) edge[line width=1pt,->,bend left=30,left] node{7} (10)
        
        (16) edge[line width=1pt,->,bend left=50,below] node{8} (15)
        (10) edge[line width=1pt,->,bend left=30,left] node{8} (5);

    \path[use as bounding box] (-2,-4.4) rectangle (6.8,0.8);
\end{tikzpicture}%
}%
        \vspace{-0.25em}\\
        (a) Expensive Qiskit routing\vspace{0.25em}
        \\
        \scalebox{1.5}{\begin{tikzpicture} [node distance=1.2cm,scale=0.6, every node/.style={scale=0.6}]
    \node[fill={rgb,190:red,72;green,132;blue,189},draw,circle,line width=0.5pt,minimum size=0.75cm,] (0) {0};
    \node[fill={rgb,190:red,112;green,169;blue,45},draw,circle,line width=0.5pt,minimum size=0.75cm,right of=0] (1) {1};
    \node[fill={rgb,190:red,180;green,56;blue,101},draw,circle,line width=0.5pt,minimum size=0.75cm,right of=1] (2) {2};
    \node[fill={rgb,190:red,112;green,169;blue,45},draw,circle,line width=0.5pt,minimum size=0.75cm,right of=2] (3) {3};
    \node[fill={rgb,190:red,72;green,132;blue,189},draw,circle,line width=0.5pt,minimum size=0.75cm,right of=3] (4) {4};
    \node[fill={rgb,190:red,72;green,132;blue,189},draw,circle,line width=0.5pt,minimum size=0.75cm,below of=0] (5) {5};
    \node[fill={rgb,190:red,180;green,56;blue,101},draw,circle,line width=0.5pt,minimum size=0.75cm,below of=1] (6) {6};
    \node[fill={rgb,190:red,72;green,132;blue,189},draw,circle,line width=0.5pt,minimum size=0.75cm,below of=2] (7) {7};
    \node[fill={rgb,190:red,72;green,132;blue,189},draw,circle,line width=0.5pt,minimum size=0.75cm,below of=3] (8) {8};
    \node[fill={rgb,190:red,72;green,132;blue,189},draw,circle,line width=0.5pt,minimum size=0.75cm,below of=4] (9) {9};
    \node[fill={rgb,190:red,72;green,132;blue,189},draw,circle,line width=0.5pt,minimum size=0.75cm,below of=5] (10) {10};
    \node[fill={rgb,190:red,72;green,132;blue,189},draw,circle,line width=0.5pt,minimum size=0.75cm,below of=6] (11) {11};
    \node[fill={rgb,190:red,72;green,132;blue,189},draw,circle,line width=0.5pt,minimum size=0.75cm,below of=7] (12) {12};
    \node[fill={rgb,190:red,72;green,132;blue,189},draw,circle,line width=0.5pt,minimum size=0.75cm,below of=8] (13) {13};
    \node[fill={rgb,190:red,72;green,132;blue,189},draw,circle,line width=0.5pt,minimum size=0.75cm,below of=9] (14) {14};
    \node[fill={rgb,190:red,72;green,132;blue,189},draw,circle,line width=0.5pt,minimum size=0.75cm,below of=10] (15) {15};
    \node[fill={rgb,190:red,72;green,132;blue,189},draw,circle,line width=0.5pt,minimum size=0.75cm,below of=11] (16) {16};
    \node[fill={rgb,190:red,72;green,132;blue,189},draw,circle,line width=0.5pt,minimum size=0.75cm,below of=12] (17) {17};
    \node[fill={rgb,190:red,72;green,132;blue,189},draw,circle,line width=0.5pt,minimum size=0.75cm,below of=13] (18) {18};
    \node[fill={rgb,190:red,180;green,56;blue,101},draw,circle,line width=0.5pt,minimum size=0.75cm,below of=14] (19) {19};

    \draw
        (0) edge[line width=2pt] node{} (1)
        (1) edge[line width=2pt] node{} (2)
        (2) edge[line width=2pt] node{} (3)
        (3) edge[line width=2pt] node{} (4)
        (0) edge[line width=2pt] node{} (5)
        (4) edge[line width=2pt] node{} (9)
        (5) edge[line width=2pt] node{} (6)
        (6) edge[line width=2pt] node{} (7)
        (7) edge[line width=2pt] node{} (8)
        (8) edge[line width=2pt] node{} (9)
        (5) edge[line width=2pt] node{} (10)
        (9) edge[line width=2pt] node{} (14)
        (7) edge[line width=2pt] node{} (12)
        (10) edge[line width=2pt] node{} (11)
        (11) edge[line width=2pt] node{} (12)
        (12) edge[line width=2pt] node{} (13)
        (13) edge[line width=2pt] node{} (14)
        (10) edge[line width=2pt] node{} (15)
        (14) edge[line width=2pt] node{} (19)
        (15) edge[line width=2pt] node{} (16)
        (16) edge[line width=2pt] node{} (17)
        (17) edge[line width=2pt] node{} (18)
        (18) edge[line width=2pt] node{} (19)
        
        (19) edge[line width=1pt,->,bend right=50,right] node{1} (14)
        (6) edge[line width=1pt,->,bend right=50,above] node{1} (5)
        
        (5) edge[line width=1pt,->,bend left=50,left] node{2} (0)
        (14) edge[line width=1pt,->,bend right=50,right] node{2} (9)
        
        (0) edge[line width=1pt,->,bend left=50,above] node{3} (1)
        (9) edge[line width=1pt,->,bend right=50,right] node{3} (4)
        
        (4) edge[line width=1pt,->,bend right=50,above] node{4} (3);

    \path[use as bounding box] (-2,-4.4) rectangle (6.8,0.8);
\end{tikzpicture}%
}%
        \vspace{-1em}\\
        (b) Efficient Trios routing
    \end{minipage}
    \caption {Example routing from Qiskit (a) vs. Trios (b) for a single Toffoli operation. Circles represent qubits and lines indicate two qubits are connected.  Input qubits are highlighted in red.
    SWAP arrows are labeled by timestep.
    The routed locations for Trios routing are highlighted in green while Qiskit moves them several times.
    Qiskit adds 16 SWAPs (=48 CNOTs), some during the Toffoli, while Trios adds only 7 SWAPs (=21 CNOTs) all before the Toffoli.
    Performing multiple passes of decomposition allows direct routing and enables this huge reduction in communication, increasing the probability of program success.
    }
    \label{fig:swap_example}
\end{figure}
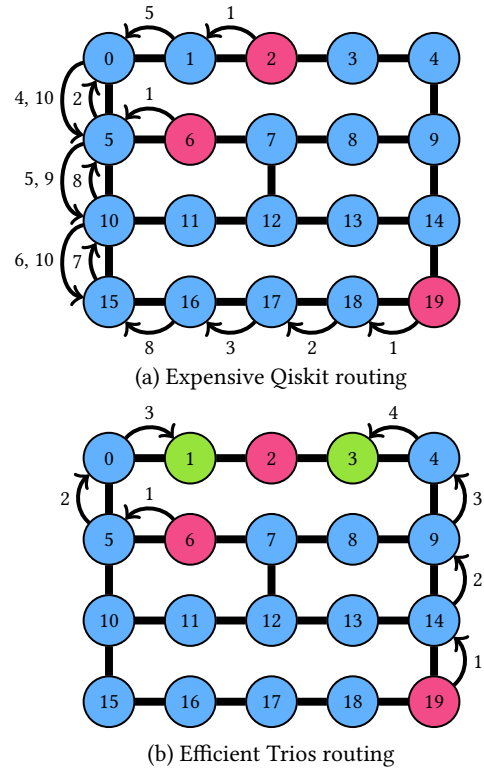

\label{sec:introduction}

In recent years, quantum hardware has improved dramatically in terms of number of accessible quantum bits (qubits), device error rates, and qubit lifetimes. However, we are still years away from obtaining fully error corrected devices which are required to run important algorithms like Grover's \cite{grover} and Shor's \cite{shor}. In the current Noisy-Intermediate Scale Quantum (NISQ) \cite{NISQ} era, where despite recent substantial improvements, error rates on current devices are still prohibitive, requiring programs to be highly optimized to have a good chance at succeeding. 

Quantum program compilation involves many passes of transformations and optimizations similar in many ways to classical compilers. Some optimizations occur at the abstract circuit level, independent of the underlying hardware, such as gate cancellation \cite{cancel}. One of the first steps usually taken is to convert an input program into a gate set (ISA) supported by the target hardware. For example, on IBM devices, gates are typically rewritten using only gates in the set $\{u1, u2, u3, cx\}$ \cite{ibmq} (single-qubit gates and the common CNOT gate described later). One critical limitation of many current available architectures is the inability to execute more complex multi-qubit operations, like the Toffoli, directly; instead, these gates must be decomposed into the supported one- and two-qubit gates. Furthermore, many current superconducting architectures only support two qubit operations on adjacent hardware qubits wired together with a coupler.  This requires the insertion of additional operations called SWAPs to move the data onto adjacent (and connected) qubits.

The process of transforming an optimized and decomposed program to the desired target is typically broken down into three distinct steps: decomposing the program into basic gates, mapping the logical qubits of a program to hardware qubits and routing interacting qubits so that they are adjacent on hardware when they interact, and scheduling operations in order to minimize total program run time (depth) or to minimize errors due to crosstalk \cite{xtalk}. Each of these steps is critical to the success of the input program. A well-mapped and well-routed program will reduce the total number of communication operations added and subsequently reduce the compiled program's depth, both of which will increase the chance of success. Conventionally, these three steps occurs sequentially. By doing so, current strategies are unable to account for structure in the input program, resulting in inefficient routing of qubits.  An optimal compiler could find the best routing despite the lack of structure but at the cost of much slower compilation.  Consider the SWAP paths inserted by IBM's Qiskit compiler for a single Toffoli compiled to IBM's Johannesburg device in Figure \ref{fig:swap_example}a. This baseline strategy adds a large number of unnecessary SWAPs as it individually routes each CNOT composing the Toffoli, dramatically reducing the probability of successful execution.

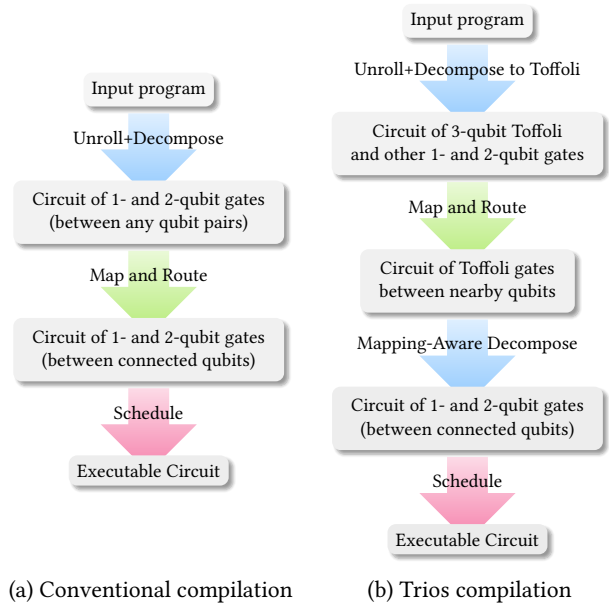
\begin{figure}
    \centering
    \begin{minipage}[][][b]{0.5\columnwidth}
        \centering
        \scalebox{0.8}{\begin{tikzpicture} [node distance=1.2cm,scale=1,inner sep=4pt]
    \node[rectangle, rounded corners, fill=black!4!white, blur shadow={shadow blur steps=5}] (0) {Input program};
    \node[arrowstyle1, below=0.1 of 0] (0a) {~~~~};
    \node[rectangle, below=0.25 of 0] (0b) {Unroll+Decompose};
    \node[rectangle, rounded corners, fill=black!5!white, blur shadow={shadow blur steps=5}, below=-0.3 of 0a] (1) {
        \begin{tabular}{c}Circuit of 1- and 2-qubit gates \\ (between any qubit pairs)\end{tabular}};
    \node[arrowstyle2, below=0.1 of 1] (1a) {~~~~};
    \node[rectangle, below=0.25 of 1] (1b) {Map and Route};
    \node[rectangle, rounded corners, fill=black!6!white, blur shadow={shadow blur steps=5}, below=-0.3 of 1a] (2) {
        \begin{tabular}{c}Circuit of 1- and 2-qubit gates \\ (between connected qubits)\end{tabular}};
    \node[arrowstyle3, below=0.1 of 2] (2a) {~~~~};
    \node[rectangle, below=0.25 of 2] (2b) {Schedule};
    \node[rectangle, rounded corners, fill=black!7!white, blur shadow={shadow blur steps=5}, below=-0.3 of 2a] (3) {Executable Circuit};
\end{tikzpicture}}%
    \end{minipage}%
    \begin{minipage}[][][b]{0.5\columnwidth}
        \centering
        \scalebox{0.8}{\begin{tikzpicture} [node distance=1.2cm,scale=1,inner sep=4pt]
    \node[rectangle, rounded corners, fill=black!4!white, blur shadow={shadow blur steps=5}] (0) {Input program};
    \node[arrowstyle1, below=0.1 of 0] (0a) {~~~~};
    \node[rectangle, below=0.25 of 0] (0b) {Unroll+Decompose to Toffoli};
    
    \node[rectangle, rounded corners, fill=black!5!white, blur shadow={shadow blur steps=5}, below=-0.3 of 0a] (1) {
        \begin{tabular}{c}Circuit of 3-qubit Toffoli \\ and other 1- and 2-qubit gates\end{tabular}};
    \node[arrowstyle2, below=0.1 of 1] (1a) {~~~~};
    \node[rectangle, below=0.25 of 1] (1b) {Map and Route};
    
    \node[rectangle, rounded corners, fill=black!6!white, blur shadow={shadow blur steps=5}, below=-0.3 of 1a] (2) {
        \begin{tabular}{c}Circuit of Toffoli gates \\ between nearby qubits\end{tabular}};
    \node[arrowstyle1, below=0.1 of 2] (2a) {~~~~};
    \node[rectangle, below=0.25 of 2] (2b) {Mapping-Aware Decompose};
    
    \node[rectangle, rounded corners, fill=black!7!white, blur shadow={shadow blur steps=5}, below=-0.3 of 2a] (3) {
        \begin{tabular}{c}Circuit of 1- and 2-qubit gates \\ (between connected qubits)\end{tabular}};
    \node[arrowstyle3, below=0.1 of 3] (3a) {~~~~};
    \node[rectangle, below=0.25 of 3] (3b) {Schedule};
    
    \node[rectangle, rounded corners, fill=black!8!white, blur shadow={shadow blur steps=5}, below=-0.3 of 3a] (4) {Executable Circuit};
\end{tikzpicture}}%
    \end{minipage}
    
    \vspace{1em}
    
    \begin{minipage}[][][b]{0.5\columnwidth}
        \centering
        (a) Conventional compilation
    \end{minipage}%
    \begin{minipage}[][][b]{0.5\columnwidth}
        \centering
        (b) Trios compilation
    \end{minipage}
    
    \caption{(a) Typical compilation passes used by Qiskit (simplified). (b) Trios compilation passes.}
    \label{fig:tool-flow}
\end{figure}

Our approach, Orchestrated Trios (Trios) decomposes and routes qubits in multiple stages, as seen in Figure \ref{fig:tool-flow}b. For example, first decompose an input program to one- two-, \textit{and} three-qubit gates (e.g. do not decompose Toffoli gates) and route as before except for three-qubits, route all three to a common location with minimal SWAPs. This new program can then undergo a second round of decomposition to produce a circuit containing only hardware permitted one- and two-qubit gates. The second round may use the now known mapping (locations of data qubits on the device) to generate fine-tuned decompositions for the architecture.

This layered approach has a major advantage over current routing techniques: we are better able to capture program structure by inspecting intermediate complex operations for routing. This better informs how qubits should be moved around the device during program execution. In Figure \ref{fig:swap_example}, the Trios strategy reduces the total number of SWAPs added to 21: fewer than half compared to Qiskit.  This was an extreme example we selected to present the issue, not an average case.

We specifically propose a two-pass approach to circuit decomposition.  We will focus on superconducting hardware systems like IBM's cloud accessible devices, but our strategy can easily be adapted to other systems. An overview of our compilation structure is found in Figure \ref{fig:tool-flow}b. This strategy has a substantial benefit on the overall success rate of programs. We demonstrate these improvements by executing Toffoli gates on a real IBM quantum computer and estimating success probability of a suite of benchmarks via simulation.

Our contributions are as follows:
\begin{itemize}
    \item A new compiler structure, Trios, with two passes for decomposition with a modified routing pass in between which greatly improves qubit routing.
    \item A simple method for architecture-tuned Toffoli decompositions during the second decompose pass that allows for a new kind of location-aware optimization.
    \item On Toffoli-only experiments, Trios reduces the total number of gates by 35\% geomean (geometric mean) resulting in 23\% geomean increase in success rate when run on real IBM hardware as compared to Qiskit.
    \item On near-term algorithms shown in Figure \ref{fig:benchmark-success-norm} (4 to 20 qubit benchmarks), Trios reduces total gate count by 37\% geomean resulting in 344\% geomean increase in (or 4.44x) simulated success rate on IBM Johannesburg with noise rates of near-future hardware as compared to programs compiled without Trios.  A sensitivity analysis over four architecture types shows the benefit range from 133\% to 3020\% increase in success rate.
\end{itemize}
\section{Background}
\label{sec:background}

\subsection{Quantum Computing Basics}
The most basic object in quantum computing is the quantum bit (qubit). Unlike a classical bit which is either 0 or 1, the qubit has two basis states $\ket{0}$ and $\ket{1}$ and can exist as a linear superposition over these two states, i.e. for a quantum state $\ket{\psi} = \alpha\ket{0} + \beta\ket{1}$ with $\alpha, \beta \in \mathbb{C}$ and $\|\alpha\|^2 + \|\beta\|^2 = 1$. In general, a quantum system consisting of $n$ qubits can exist in a linear superposition of $2^n$ basis states in contrast to a classical system of $n$ bits which can exist as exactly a single of these states. An important feature which gives quantum computing its power is the ability to \textit{entangle} qubits via two qubit operations like the CNOT. This, along with quantum interference between the complex amplitudes, allows quantum programs to solve certain problems faster than classical computers.

While a qubit system can exist in these superpositions during computation, at the end of the computation, the qubits are measured producing a classical binary outcome. The probability of each outcome depends on the amplitude of each basis state (the values of $\alpha, \beta, \gamma, \dots$). Consequently, since the outcome of a quantum program is a classical bitstring and because quantum systems are inherently noisy, programs are usually run thousands of times to obtain a distribution over possible answers. A comprehensive background can be found in \cite{mikeike}. 

\subsection{Quantum Circuits}

Quantum programs are typically represented as a circuit which, like a classical program, is an ordered list of instructions. Here the instructions are quantum logic gates applied to qubits. The input circuit may not be expressed in the instruction set supported by the underlying hardware or it might even be structured as hierarchical modules.

Quantum circuits have a single line for each qubit, with time flowing from left to right. Gates in a quantum circuits have the same number of inputs and outputs and gates on disjoint sets of lines can be executed in parallel. Single qubit gates are represented as a box labeled with the indicated operation and controlled operations, like the CNOT and Toffoli, have one or two controls respectively indicated by $\bullet$ and target given by $\oplus$.

\begin{figure}
    \centering
    \begin{minipage}[][][b]{\columnwidth}
    \centering
    \scalebox{1}{%
        \input{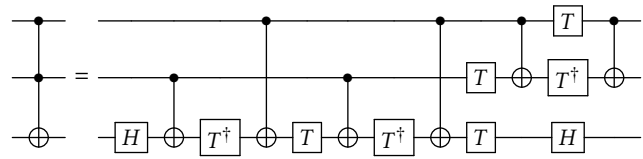}%
    }%
    \caption{A 6-CNOT decomposition of the Toffoli gate.}
    \label{fig:6-cnot-decomp}
    \end{minipage}\\
\end{figure}

Currently available superconducting quantum hardware, like that of IBM and Rigetti, only supports one-qubit gates and two-qubit gates on specific pairs. Therefore, more complex instructions must be decomposed into multiple simpler, supported operations. For example, many quantum algorithms and subroutines make use of the Toffoli gate, a three-input gate which performs the logical AND between two controls bits and writes the output onto the target bit. This gate cannot be executed directly on available hardware and instead is decomposed into an equivalent sequence of one- and two-qubit operations. Two such popular decompositions are given in Figures \ref{fig:6-cnot-decomp}, \ref{fig:8-cnot-decomp}. There are two key distinctions in these decompositions illustrating a more general trade off. The first \cite{mikeike} is the most popular decomposition using only 6 CNOT gates but requires CNOTs between all three pairs of qubits.  This would require inserted SWAPs or a device connectivity containing a triangle. The second \cite{craig8} uses a total of 8 CNOT gates and requires all three inputs be only linearly connected (only two of the three qubit pairs are required to be connected). While the first is apparently more efficient, this is not true if the connectivity of the underlying hardware does not directly support it.  It is more efficient to use the 8-CNOT version than use the 6-CNOT version with added SWAPs.

For superconducting qubits, current quantum computers supports gates only between adjacent hardware qubits. In order to use qubits which are currently mapped far apart on the hardware, extra SWAP operations must be inserted, each of these SWAPs is usually decomposed as a series of 3 CNOT gates (equivalent to a classical memory in-place swap using 3 alternating XORs). In the case of the 6-CNOT Toffoli decomposition above, when mapped to a device with linear or square grid connectivity, no triangles exist so extra SWAPs will need to be inserted, resulting in a greater total number of CNOTs due to the mismatch with hardware details.

\begin{figure}
    \begin{minipage}[][][b]{\columnwidth}
    \centering%
    \scalebox{1}{%
        \input{figs/8cnot-decomp}%
    }%
    \caption{An 8-CNOT decomposition of the Toffoli gate.}
    \label{fig:8-cnot-decomp}
    \end{minipage}
\end{figure}

\subsection{Current Quantum Devices}

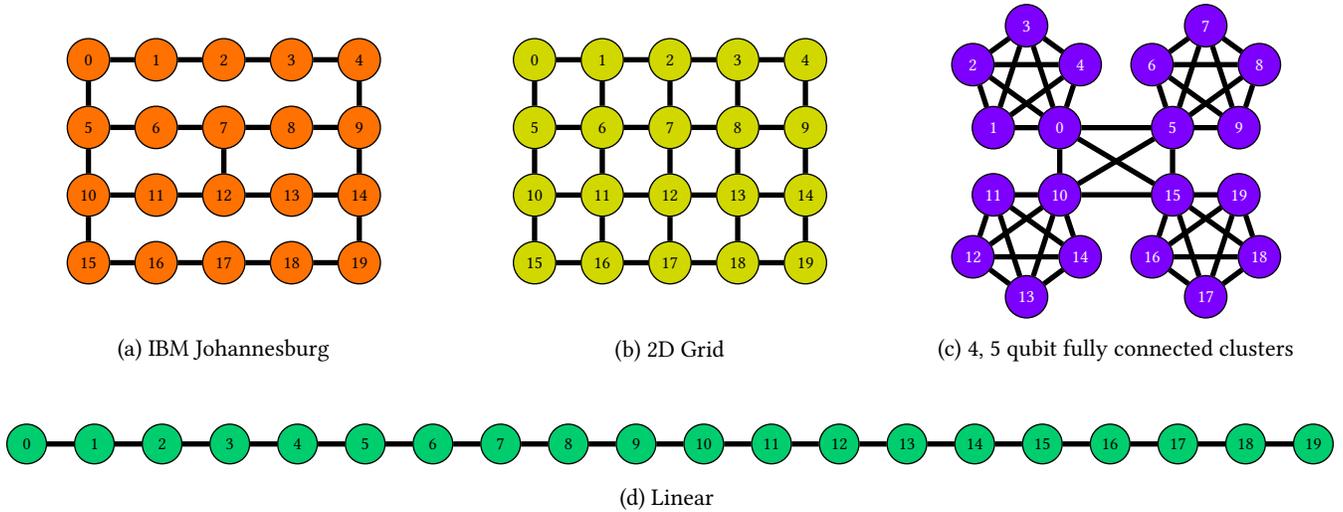
\begin{figure*}
    \centering
    \hfill%
    \begin{minipage}[][][b]{0.3333\textwidth}
        \centering
        \begin{tikzpicture} [node distance=1.2cm,scale=0.75, every node/.style={scale=0.75}]
        \node[fill={rgb,255:red,255;green,113;blue,0},draw,circle,line width=0.5pt,minimum size=0.75cm,] (0) {0};
        \node[fill={rgb,255:red,255;green,113;blue,0},draw,circle,line width=0.5pt,minimum size=0.75cm,right of=0] (1) {1};
        \node[fill={rgb,255:red,255;green,113;blue,0},draw,circle,line width=0.5pt,minimum size=0.75cm,right of=1] (2) {2};
        \node[fill={rgb,255:red,255;green,113;blue,0},draw,circle,line width=0.5pt,minimum size=0.75cm,right of=2] (3) {3};
        \node[fill={rgb,255:red,255;green,113;blue,0},draw,circle,line width=0.5pt,minimum size=0.75cm,right of=3] (4) {4};
        \node[fill={rgb,255:red,255;green,113;blue,0},draw,circle,line width=0.5pt,minimum size=0.75cm,below of=0] (5) {5};
        \node[fill={rgb,255:red,255;green,113;blue,0},draw,circle,line width=0.5pt,minimum size=0.75cm,below of=1] (6) {6};
        \node[fill={rgb,255:red,255;green,113;blue,0},draw,circle,line width=0.5pt,minimum size=0.75cm,below of=2] (7) {7};
        \node[fill={rgb,255:red,255;green,113;blue,0},draw,circle,line width=0.5pt,minimum size=0.75cm,below of=3] (8) {8};
        \node[fill={rgb,255:red,255;green,113;blue,0},draw,circle,line width=0.5pt,minimum size=0.75cm,below of=4] (9) {9};
        \node[fill={rgb,255:red,255;green,113;blue,0},draw,circle,line width=0.5pt,minimum size=0.75cm,below of=5] (10) {10};
        \node[fill={rgb,255:red,255;green,113;blue,0},draw,circle,line width=0.5pt,minimum size=0.75cm,below of=6] (11) {11};
        \node[fill={rgb,255:red,255;green,113;blue,0},draw,circle,line width=0.5pt,minimum size=0.75cm,below of=7] (12) {12};
        \node[fill={rgb,255:red,255;green,113;blue,0},draw,circle,line width=0.5pt,minimum size=0.75cm,below of=8] (13) {13};
        \node[fill={rgb,255:red,255;green,113;blue,0},draw,circle,line width=0.5pt,minimum size=0.75cm,below of=9] (14) {14};
        \node[fill={rgb,255:red,255;green,113;blue,0},draw,circle,line width=0.5pt,minimum size=0.75cm,below of=10] (15) {15};
        \node[fill={rgb,255:red,255;green,113;blue,0},draw,circle,line width=0.5pt,minimum size=0.75cm,below of=11] (16) {16};
        \node[fill={rgb,255:red,255;green,113;blue,0},draw,circle,line width=0.5pt,minimum size=0.75cm,below of=12] (17) {17};
        \node[fill={rgb,255:red,255;green,113;blue,0},draw,circle,line width=0.5pt,minimum size=0.75cm,below of=13] (18) {18};
        \node[fill={rgb,255:red,255;green,113;blue,0},draw,circle,line width=0.5pt,minimum size=0.75cm,below of=14] (19) {19};
    
        \draw
            (0) edge[line width=2pt] node{} (1)
            (1) edge[line width=2pt] node{} (2)
            (2) edge[line width=2pt] node{} (3)
            (3) edge[line width=2pt] node{} (4)
            (0) edge[line width=2pt] node{} (5)
            (4) edge[line width=2pt] node{} (9)
            (5) edge[line width=2pt] node{} (6)
            (6) edge[line width=2pt] node{} (7)
            (7) edge[line width=2pt] node{} (8)
            (8) edge[line width=2pt] node{} (9)
            (5) edge[line width=2pt] node{} (10)
            (9) edge[line width=2pt] node{} (14)
            (7) edge[line width=2pt] node{} (12)
            (10) edge[line width=2pt] node{} (11)
            (11) edge[line width=2pt] node{} (12)
            (12) edge[line width=2pt] node{} (13)
            (13) edge[line width=2pt] node{} (14)
            (10) edge[line width=2pt] node{} (15)
            (14) edge[line width=2pt] node{} (19)
            (15) edge[line width=2pt] node{} (16)
            (16) edge[line width=2pt] node{} (17)
            (17) edge[line width=2pt] node{} (18)
            (18) edge[line width=2pt] node{} (19);
\end{tikzpicture}%
    \end{minipage}\hfill%
    \begin{minipage}[][][b]{0.3333\textwidth}
        \centering
        \begin{tikzpicture} [node distance=1.2cm,scale=0.75, every node/.style={scale=0.75}]
        \node[fill={rgb,255:red,209;green,216;blue,0},draw,circle,line width=0.5pt,minimum size=0.75cm,] (0) {0};
        \node[fill={rgb,255:red,209;green,216;blue,0},draw,circle,line width=0.5pt,minimum size=0.75cm,right of=0] (1) {1};
        \node[fill={rgb,255:red,209;green,216;blue,0},draw,circle,line width=0.5pt,minimum size=0.75cm,right of=1] (2) {2};
        \node[fill={rgb,255:red,209;green,216;blue,0},draw,circle,line width=0.5pt,minimum size=0.75cm,right of=2] (3) {3};
        \node[fill={rgb,255:red,209;green,216;blue,0},draw,circle,line width=0.5pt,minimum size=0.75cm,right of=3] (4) {4};
        \node[fill={rgb,255:red,209;green,216;blue,0},draw,circle,line width=0.5pt,minimum size=0.75cm,below of=0] (5) {5};
        \node[fill={rgb,255:red,209;green,216;blue,0},draw,circle,line width=0.5pt,minimum size=0.75cm,below of=1] (6) {6};
        \node[fill={rgb,255:red,209;green,216;blue,0},draw,circle,line width=0.5pt,minimum size=0.75cm,below of=2] (7) {7};
        \node[fill={rgb,255:red,209;green,216;blue,0},draw,circle,line width=0.5pt,minimum size=0.75cm,below of=3] (8) {8};
        \node[fill={rgb,255:red,209;green,216;blue,0},draw,circle,line width=0.5pt,minimum size=0.75cm,below of=4] (9) {9};
        \node[fill={rgb,255:red,209;green,216;blue,0},draw,circle,line width=0.5pt,minimum size=0.75cm,below of=5] (10) {10};
        \node[fill={rgb,255:red,209;green,216;blue,0},draw,circle,line width=0.5pt,minimum size=0.75cm,below of=6] (11) {11};
        \node[fill={rgb,255:red,209;green,216;blue,0},draw,circle,line width=0.5pt,minimum size=0.75cm,below of=7] (12) {12};
        \node[fill={rgb,255:red,209;green,216;blue,0},draw,circle,line width=0.5pt,minimum size=0.75cm,below of=8] (13) {13};
        \node[fill={rgb,255:red,209;green,216;blue,0},draw,circle,line width=0.5pt,minimum size=0.75cm,below of=9] (14) {14};
        \node[fill={rgb,255:red,209;green,216;blue,0},draw,circle,line width=0.5pt,minimum size=0.75cm,below of=10] (15) {15};
        \node[fill={rgb,255:red,209;green,216;blue,0},draw,circle,line width=0.5pt,minimum size=0.75cm,below of=11] (16) {16};
        \node[fill={rgb,255:red,209;green,216;blue,0},draw,circle,line width=0.5pt,minimum size=0.75cm,below of=12] (17) {17};
        \node[fill={rgb,255:red,209;green,216;blue,0},draw,circle,line width=0.5pt,minimum size=0.75cm,below of=13] (18) {18};
        \node[fill={rgb,255:red,209;green,216;blue,0},draw,circle,line width=0.5pt,minimum size=0.75cm,below of=14] (19) {19};
    
        \draw
            (0) edge[line width=2pt] node{} (1)
            (1) edge[line width=2pt] node{} (2)
            (2) edge[line width=2pt] node{} (3)
            (3) edge[line width=2pt] node{} (4)
            (5) edge[line width=2pt] node{} (6)
            (6) edge[line width=2pt] node{} (7)
            (7) edge[line width=2pt] node{} (8)
            (8) edge[line width=2pt] node{} (9)
            (10) edge[line width=2pt] node{} (11)
            (11) edge[line width=2pt] node{} (12)
            (12) edge[line width=2pt] node{} (13)
            (13) edge[line width=2pt] node{} (14)
            (15) edge[line width=2pt] node{} (16)
            (16) edge[line width=2pt] node{} (17)
            (17) edge[line width=2pt] node{} (18)
            (18) edge[line width=2pt] node{} (19)
            (0) edge[line width=2pt] node{} (5)
            (1) edge[line width=2pt] node{} (6)
            (2) edge[line width=2pt] node{} (7)
            (3) edge[line width=2pt] node{} (8)
            (4) edge[line width=2pt] node{} (9)
            (5) edge[line width=2pt] node{} (10)
            (6) edge[line width=2pt] node{} (11)
            (7) edge[line width=2pt] node{} (12)
            (8) edge[line width=2pt] node{} (13)
            (9) edge[line width=2pt] node{} (14)
            (10) edge[line width=2pt] node{} (15)
            (11) edge[line width=2pt] node{} (16)
            (12) edge[line width=2pt] node{} (17)
            (13) edge[line width=2pt] node{} (18)
            (14) edge[line width=2pt] node{} (19);
\end{tikzpicture}%
    \end{minipage}\hfill%
    \begin{minipage}[][][b]{0.3333\textwidth}
        \centering
        \begin{tikzpicture} [node distance=1.2cm,scale=0.75, every node/.style={scale=0.75}]
        \node[fill={rgb,255:red,125;green,0;blue,255},text=white,draw,circle,line width=0.5pt,minimum size=0.75cm] (0) at (0.588,-0.809) {0};
        \node[fill={rgb,255:red,125;green,0;blue,255},text=white,draw,circle,line width=0.5pt,minimum size=0.75cm] (1) at (-0.588,-0.809) {1};
        \node[fill={rgb,255:red,125;green,0;blue,255},text=white,draw,circle,line width=0.5pt,minimum size=0.75cm] (2) at (-0.951,0.309) {2};
        \node[fill={rgb,255:red,125;green,0;blue,255},text=white,draw,circle,line width=0.5pt,minimum size=0.75cm] (3) at (0, 1) {3};
        \node[fill={rgb,255:red,125;green,0;blue,255},text=white,draw,circle,line width=0.5pt,minimum size=0.75cm] (4) at (0.951, 0.309) {4};
        \node[fill={rgb,255:red,125;green,0;blue,255},text=white,draw,circle,line width=0.5pt,minimum size=0.75cm] (5) at (2.588,-0.809) {5};
        \node[fill={rgb,255:red,125;green,0;blue,255},text=white,draw,circle,line width=0.5pt,minimum size=0.75cm] (6) at (2.225,0.309) {6};
        \node[fill={rgb,255:red,125;green,0;blue,255},text=white,draw,circle,line width=0.5pt,minimum size=0.75cm] (7) at (3.176,1) {7};
        \node[fill={rgb,255:red,125;green,0;blue,255},text=white,draw,circle,line width=0.5pt,minimum size=0.75cm] (8) at (4.127,0.309) {8};
        \node[fill={rgb,255:red,125;green,0;blue,255},text=white,draw,circle,line width=0.5pt,minimum size=0.75cm] (9) at (3.764,-0.809) {9};
        \node[fill={rgb,255:red,125;green,0;blue,255},text=white,draw,circle,line width=0.5pt,minimum size=0.75cm] (10) at (0.588,-2) {10};
        \node[fill={rgb,255:red,125;green,0;blue,255},text=white,draw,circle,line width=0.5pt,minimum size=0.75cm] (11) at (-0.588,-2) {11};
        \node[fill={rgb,255:red,125;green,0;blue,255},text=white,draw,circle,line width=0.5pt,minimum size=0.75cm] (12) at (-0.951,-3.1) {12};
        \node[fill={rgb,255:red,125;green,0;blue,255},text=white,draw,circle,line width=0.5pt,minimum size=0.75cm] (13) at (0,-3.809) {13};
        \node[fill={rgb,255:red,125;green,0;blue,255},text=white,draw,circle,line width=0.5pt,minimum size=0.75cm] (14) at (0.951,-3.1) {14};
        \node[fill={rgb,255:red,125;green,0;blue,255},text=white,draw,circle,line width=0.5pt,minimum size=0.75cm] (15) at (2.588,-2) {15};
        \node[fill={rgb,255:red,125;green,0;blue,255},text=white,draw,circle,line width=0.5pt,minimum size=0.75cm] (16) at (2.225,-3.1) {16};
        \node[fill={rgb,255:red,125;green,0;blue,255},text=white,draw,circle,line width=0.5pt,minimum size=0.75cm] (17) at (3.176,-3.809) {17};
        \node[fill={rgb,255:red,125;green,0;blue,255},text=white,draw,circle,line width=0.5pt,minimum size=0.75cm] (18) at (4.127,-3.1) {18};
        \node[fill={rgb,255:red,125;green,0;blue,255},text=white,draw,circle,line width=0.5pt,minimum size=0.75cm] (19) at (3.764,-2) {19};
        \draw
        (0) edge[line width=2pt] node{} (1)
        (0) edge[line width=2pt] node{} (2)
        (0) edge[line width=2pt] node{} (3)
        (0) edge[line width=2pt] node{} (4)
        (1) edge[line width=2pt] node{} (2)
        (1) edge[line width=2pt] node{} (3)
        (1) edge[line width=2pt] node{} (4)
        (2) edge[line width=2pt] node{} (3)
        (2) edge[line width=2pt] node{} (4)
        (3) edge[line width=2pt] node{} (4)
        (0) edge[line width=2pt] node{} (5)
        (0) edge[line width=2pt] node{} (10)
        
        (0) edge[line width=2pt] node{} (15)
        (10) edge[line width=2pt] node{} (5)
        
        (5) edge[line width=2pt] node{} (6)
        (5) edge[line width=2pt] node{} (7)
        (5) edge[line width=2pt] node{} (8)
        (5) edge[line width=2pt] node{} (9)
        (6) edge[line width=2pt] node{} (7)
        (6) edge[line width=2pt] node{} (8)
        (6) edge[line width=2pt] node{} (9)
        (7) edge[line width=2pt] node{} (8)
        (7) edge[line width=2pt] node{} (9)
        (8) edge[line width=2pt] node{} (9)
        (5) edge[line width=2pt] node{} (15)
        
        (10) edge[line width=2pt] node{} (11)
        (10) edge[line width=2pt] node{} (12)
        (10) edge[line width=2pt] node{} (13)
        (10) edge[line width=2pt] node{} (14)
        (11) edge[line width=2pt] node{} (12)
        (11) edge[line width=2pt] node{} (13)
        (11) edge[line width=2pt] node{} (14)
        (12) edge[line width=2pt] node{} (13)
        (12) edge[line width=2pt] node{} (14)
        (13) edge[line width=2pt] node{} (14)
        (10) edge[line width=2pt] node{} (15)
        
        (15) edge[line width=2pt] node{} (16)
        (15) edge[line width=2pt] node{} (17)
        (15) edge[line width=2pt] node{} (18)
        (15) edge[line width=2pt] node{} (19)
        (16) edge[line width=2pt] node{} (17)
        (16) edge[line width=2pt] node{} (18)
        (16) edge[line width=2pt] node{} (19)
        (17) edge[line width=2pt] node{} (18)
        (17) edge[line width=2pt] node{} (19)
        (18) edge[line width=2pt] node{} (19);
\end{tikzpicture}%
    \end{minipage}\hfill%
    \vspace{0.75em}\\
    \begin{minipage}[][][b]{0.3333\textwidth}
        \centering
        (a) IBM Johannesburg
    \end{minipage}\hfill%
    \begin{minipage}[][][b]{0.3333\textwidth}
        \centering
        (b) 2D Grid
    \end{minipage}\hfill%
    \begin{minipage}[][][b]{0.3333\textwidth}
        \centering
        (c) 4, 5 qubit fully connected clusters
    \end{minipage}\hfill%
    \vspace{1.5em}\\
    \begin{minipage}[][][b]{\textwidth}
        \centering
        \vspace{1em}
        \begin{tikzpicture} [node distance=1.2cm,scale=0.75, every node/.style={scale=0.75}]
        \node[fill={rgb,255:red,0;green,205;blue,110},draw,line width=0.5pt,circle,minimum size=20 pt,] (0) {0};
        \node[fill={rgb,255:red,0;green,205;blue,110},draw,line width=0.5pt,circle,minimum size=20 pt,right of=0] (1) {1};
        \node[fill={rgb,255:red,0;green,205;blue,110},draw,line width=0.5pt,circle,minimum size=20 pt,right of=1] (2) {2};
        \node[fill={rgb,255:red,0;green,205;blue,110},draw,line width=0.5pt,circle,minimum size=20 pt,right of=2] (3) {3};
        \node[fill={rgb,255:red,0;green,205;blue,110},draw,line width=0.5pt,circle,minimum size=20 pt,right of=3] (4) {4};
        \node[fill={rgb,255:red,0;green,205;blue,110},draw,line width=0.5pt,circle,minimum size=20 pt,right of=4] (5) {5};
        \node[fill={rgb,255:red,0;green,205;blue,110},draw,line width=0.5pt,circle,minimum size=20 pt,right of=5] (6) {6};
        \node[fill={rgb,255:red,0;green,205;blue,110},draw,line width=0.5pt,circle,minimum size=20 pt,right of=6] (7) {7};
        \node[fill={rgb,255:red,0;green,205;blue,110},draw,line width=0.5pt,circle,minimum size=20 pt,right of=7] (8) {8};
        \node[fill={rgb,255:red,0;green,205;blue,110},draw,line width=0.5pt,circle,minimum size=20 pt,right of=8] (9) {9};
        \node[fill={rgb,255:red,0;green,205;blue,110},draw,line width=0.5pt,circle,minimum size=20 pt,right of=9] (10) {10};
        \node[fill={rgb,255:red,0;green,205;blue,110},draw,line width=0.5pt,circle,minimum size=20 pt,right of=10] (11) {11};
        \node[fill={rgb,255:red,0;green,205;blue,110},draw,line width=0.5pt,circle,minimum size=20 pt,right of=11] (12) {12};
        \node[fill={rgb,255:red,0;green,205;blue,110},draw,line width=0.5pt,circle,minimum size=20 pt,right of=12] (13) {13};
        \node[fill={rgb,255:red,0;green,205;blue,110},draw,line width=0.5pt,circle,minimum size=20 pt,right of=13] (14) {14};
        \node[fill={rgb,255:red,0;green,205;blue,110},draw,line width=0.5pt,circle,minimum size=20 pt,right of=14] (15) {15};
        \node[fill={rgb,255:red,0;green,205;blue,110},draw,line width=0.5pt,circle,minimum size=20 pt,right of=15] (16) {16};
        \node[fill={rgb,255:red,0;green,205;blue,110},draw,line width=0.5pt,circle,minimum size=20 pt,right of=16] (17) {17};
        \node[fill={rgb,255:red,0;green,205;blue,110},draw,line width=0.5pt,circle,minimum size=20 pt,right of=17] (18) {18};
        \node[fill={rgb,255:red,0;green,205;blue,110},draw,line width=0.5pt,circle,minimum size=20 pt,right of=18] (19) {19};
    
        \draw
            (0) edge[line width=2pt] node{} (1)
            (1) edge[line width=2pt] node{} (2)
            (2) edge[line width=2pt] node{} (3)
            (3) edge[line width=2pt] node{} (4)
            (4) edge[line width=2pt] node{} (5)
            (5) edge[line width=2pt] node{} (6)
            (6) edge[line width=2pt] node{} (7)
            (7) edge[line width=2pt] node{} (8)
            (8) edge[line width=2pt] node{} (9)
            (9) edge[line width=2pt] node{} (10)
            (10) edge[line width=2pt] node{} (11)
            (11) edge[line width=2pt] node{} (12)
            (12) edge[line width=2pt] node{} (13)
            (13) edge[line width=2pt] node{} (14)
            (14) edge[line width=2pt] node{} (15)
            (15) edge[line width=2pt] node{} (16)
            (16) edge[line width=2pt] node{} (17)
            (17) edge[line width=2pt] node{} (18)
            (18) edge[line width=2pt] node{} (19);
\end{tikzpicture}%
    \end{minipage}
    \vspace{0.25em}\\
    \begin{minipage}[][][b]{\textwidth}
        \centering
        \vspace{0.5em}
        (d) Linear
        \vspace{0.5em}
    \end{minipage}

    \caption{Example topologies of near-term quantum devices. Orange (a): IBM Johannesburg. Yellow (b): 2D Grid. Purple (c): four groups of five fully connected clusters. Green (d) Linear. Our real experiments run on Johannesburg and our simulations explore all of these topologies. Colors correspond with the bars in Figures \ref{fig:benchmark-success}, \ref{fig:benchmark-gate-ratio}, \ref{fig:benchmark-success-norm}.}
    \label{fig:device-diagrams}
\end{figure*}

\label{background:devices}
In this paper we focus primarily on currently available superconducting quantum devices. This type of hardware is the primary focus of many industry players like IBM, Rigetti, and Google \cite{rigetti, ibmq, bristlecone}. We show some representative topologies for superconducting devices in Figure \ref{fig:device-diagrams}abd.  For completeness, we include a clustered device shown in Figure \ref{fig:device-diagrams}c representative of a QCCD ion trap device such as \cite{honeywell}. These systems exhibit all of the properties previously discussed. They have a small universal supported gate set which all programs must be transformed into and only support local two-qubit operations. The connectivity of these devices is given as a \textit{coupling graph} specifying which pairs of qubits can execute CNOTs.

Furthermore, these systems are subject to a wide variety of noise which cause programs to fail. Some noise is due to manufacturing imperfections or calibration error. Some is inherent to quantum program execution resulting from the imperfect physical isolation of the qubits from the environment required to manipulate the quantum state \cite{sc_errors}. In IBM machines, the experimental devices of this work, single qubit gate errors are small, occurring on average 1 in 2000 operations. CNOT gate errors are more significant occurring on roughly 1 in 100 gates. Measurement error is also significant, with errors on the same order of magnitude as CNOT gates. Finally, qubit lifetimes (coherence times) are relatively short, allowing on the order of 50 CNOT gate durations before the qubit state is lost \cite{ibm_errors} (but gates can often run in parallel while imposing additional crosstalk error). Therefore, quantum compilation is essential to reduce both of these sources of error: add as few extra gates as possible and minimize total execution time. 

\subsection{The Compilation Problem}
\label{background:compilation}
In the NISQ era, quantum programs are highly optimized in order to reduce the effect on errors and maximize the probability the correct answer is observed. Similar to many classical programs, compilation uses a pass structure, where a set of transformation and optimizations are applied in a fixed order resulting in the compilation of an input quantum program to an executable for the target hardware \cite{scaffcc, cancel}. For the most part, these optimizations take place at the circuit-gate level. Some optimizations are hardware independent, for example, reducing total number of gates via commutativity-aware gate cancellation or find and replace with circuit identities. Other passes are focused on decomposing gates into the hardware's ISA \cite{opt1, opt2, opt3}.

One of the most important parts of this compilation process is mapping and routing the optimized program to one executable on the target hardware, typically done post-decomposition.  Quantum mechanics imposes new constraints on these than classical compilation or logic synthesis.  By the no cloning theorem, quantum states cannot be copied, only entangled, which prevents fan-out or fan-in.  Instead, the data must be routed sequentially (i.e. swapped with SWAP gates) to each place it is needed.

Compilation involves three main steps. First, mapping program qubits to hardware qubits in order to minimize the total distance between qubits that will need to be close by in the future \cite{map1, map2, map3}. Second, routing pairs of CNOT inputs to be adjacent by inserting SWAPs \cite{routing1, routing2}. Finally, scheduling operations to minimize total execution time \cite{scheduling1, xtalk}. In general, the compilation problem is computationally hard and while some attempts at optimal solutions have been pursued \cite{tan2020optimal, siraichi2018qubit, wille2014optimal} the dominant approach is heuristics. In this work we focus on two pieces of this compilation problem: decomposition and routing.

IBM's Qiskit compiler, the standard for compiling programs to execute on an IBM device, has a default sequence of passes. First, all high level optimization and analysis passes are performed and all gates are unrolled and decomposed to the target gate set. Then single passes of mapping, routing, and scheduling are performed \cite{qiskit}. 

\subsection{Evaluation Metrics}
\label{background:metrics}

When evaluating compiler methods, we use a few metrics to compare our results.  Our primary metric is program success rate, the fraction of circuit executions that result in the correct output.  Others use fidelity which can stand-in for success rate when evaluating sub-circuits where the output is not measured.  When executing a quantum algorithm, the corresponding quantum circuit is typically executed thousands of times to gather output statistics or identify the error-free result.

Program success rate is highly dependent to the noise characteristics of the quantum computer the program runs on.  The rates of these device errors can fluctuate day-to-day so we also use the simpler metric of two-qubit gate count.  The number of two-qubit operations in the final compiled circuit is inversely correlated with the success rate because they are usually the largest source of noise.

\subsection{Simulation}
\label{background:simulation} 
Simulating general quantum systems is exponentially expensive in the size of the system and therefore it is difficult to realistically model all of the errors during the execution of a quantum program. We use a simplified model for simulation to predict, specifically obtain a close upper bound on, the success rate of a program with specified gate error rates and qubit coherence times. In our simplified model, we compute the probability of a program succeeding as the probability that no gate errors occur $(p_{gate})^{n_{gates}}$ times the probability no coherence errors occur $p_{coherence}$, where the latter is computed as $e^{\Delta / T_1 + \Delta / T_2}$, where $\Delta$ is the total program duration and $T_1$ and $T_2$ are the relaxation and dephasing times, collectively decoherence.

Current error rates, while rapidly improving, are still insufficient to obtain high probabilities of success, making it difficult to compare our mid-size benchmarks that are large enough to need many SWAPs. For our simulations we use error rates 20x improved over current IBM Johannesburg error rates to obtain reasonable success rates and we study sensitivity to this choice later.

\section{Motivation: Conventional Compilation}
In this section we motivate the need for a split decomposition pass with routing in between. We look closely at the Qiskit compiler which does not effectively account for the structure in programs.  It often produces circuits with an excessive number of swaps suggesting room for improvement.%

\hide{

}

The default compilation framework in Qiskit used to transform input circuits to be executed on their hardware ensures a fully decomposed circuit before mapping, routing, and scheduling occur. As a simple example, consider three qubits placed fairly distant on IBM's Johannesburg device but for which we need to execute a Toffoli gate on them; as in Figure \ref{fig:swap_example}a. Qiskit decomposes this Toffoli as in Figure \ref{fig:6-cnot-decomp} with 6 CNOTs. Each CNOT acts on distant qubits so the many SWAPs inserted for all 6 CNOTs gets expensive quickly. When routing, we first SWAP the first interacting pair together (usually by adding SWAPs from control to target or the reverse, but a meet-in-the-middle strategy is also possible) and the qubit mapping is updated. The next CNOT is also distant so we add SWAPs to move them together and there is an even chance that the SWAPs for the second CNOT separate the two qubits that were just brought together.

Ideally, we move the third qubit to the already adjacent pair, but Qiskit cannot recognize this situation and could just as well move the other way. This is clearly sub-optimal and could continue on for the other four SWAPs. Even in the case where it makes the correct decision to move the distant third qubit, there are problems. Because pair of qubits needs to interact we may need single additional SWAPs as the qubits compete to be neighbors.  This causes the 6-CNOT Toffoli decomposition to use many more than 6 CNOTs when there is not a triangle in the qubit connectivity graph. The core idea is that the routing strategy fails to take advantage of two things. First, it has effectively forgotten the desired operation is a Toffoli which will require all three qubits be adjacent and second that a more efficient Toffoli decomposition could be chosen which is more suitable for the underlying device architecture. In the example, inefficient compilation adds a total of 16 SWAPS or 48 CNOTs in total.

Some approaches in the past have attempted to solve the first of these problems, for example by using lookahead when choosing routing strategies \cite{look1, look2} and while this helps to treat the symptoms of pre-decomposing all operations it does not remedy the underlying problem. 

\section{Orchestrated Trios}
\label{orechestra}
In this section we describe our proposed compilation structure compared to the conventional one as outlined in Figure \ref{fig:tool-flow}. Specifically, we focus on improving the routing and decomposition stages of compilation. Previously, we identified a key problem in current methods: decomposing the program to one- and two-qubit gates up front hinders the ability of heuristic-based compilers to effectively minimize the communication cost, i.e. the number of SWAPs added, and eliminates the possibility of location-aware decompositions.

We propose a new pass structure. Rather than performing a single round of decomposition and routing, we propose a split approach. Any program processing prior to decomposition stays the same.  The decomposition pass is then divided so the majority of decomposition occurs next but any Toffoli gates are left as-is before moving on to mapping and routing.

The mapping and routing passes come next like normal but must be modified slightly to handle three-qubit gates.  The mapper can simply treat the non-decomposed Toffoli as it would the equivalent 6 CNOTs for the purposes of determining which qubits most need to be placed nearby.
We then do the modified routing pass, moving \textit{groups} of qubits together instead of only pairs where all or all-but-one of the group are moved into a single neighborhood via SWAPs.  This greatly improves the effectiveness of the routing heuristics when applied to this modified routing pass.  There are some subtleties when coordinating the routing of multiple qubits to the same place to ensure the paths don't overlap.  For the purposes of our evaluations we do the following but many similar heuristic strategies are possible.

Taking the next operation to apply, we first find the shortest paths (using any shortest path algorithm on a graph) between all the pairs of qubits.  We choose the qubit with the shortest sum of paths to the other two qubits as the destination.  SWAPS following these two paths are then inserted into the circuit.  The two shortest paths are checked for overlap.  If the ending points overlap, the second is only routed to the penultimate hardware location along the swap path and the first becomes the middle qubit adjacent to both others.  This can save one valuable SWAP but doesn't affect the correctness. Once they are adjacent, the Toffoli gate is now on adjacent qubits and routing can continue to the next operation.

Finally, the second decomposition pass is run.  This is different from normal decomposition as there are only Toffoli gates to decompose and they are already mapped to neighboring qubits.  We could use the default 6-CNOT decomposition and still get the above benefit of improved routing but now that we have more information, this can be exploited to further reduce SWAPs due to a mismatch between the decomposition and the hardware connectivity.  If all three pairs of qubits are connected, then the 6-CNOT Toffoli of Figure \ref{fig:6-cnot-decomp} is best, otherwise use the 8-CNOT Toffoli of Figure \ref{fig:8-cnot-decomp}, ensuring the middle qubit is used for the middle of the decomposition (Any of the three qubits can be the target by simply moving the two H gates to that qubit).

When routing complex operations like the Toffoli, we recognize the underlying hardware does not usually support triangles in the connectivity graph but linear connectivity is sufficient for a decent decomposition.  Since we are creating operations on three qubits, the qubits must be routed into a valid linear connectivity.  That is, a configuration where each qubit is connected with at least one of the other qubits.

This method can be easily extended to be noise-aware like previous work \cite{map1, map2} by using a noise-aware mapper with the simple modification described earlier where the path-finding graph has weighed edges with the $\text{--}\log$ value of the CNOT success rate. The path distance represents the $\text{--}\log$ probability of success of that particular path where lower values indicate a higher success rate and the shortest path can be found just as before and the routing steps are unchanged.  Any routing strategy designed for one and two-qubit gates can be modified to work for one, two, and three-qubit gates and used as the first routing step of Trios.

In programs where there are no three qubit gates as in the typical NISQ benchmark, Bernstein-Vazirani \cite{bv}, which is specified directly as CNOT gates, our strategy will have no effect. Many benchmarks, however, are written using Toffoli gates because they are the quantum analog the AND gate ubiquitous in arithmetics and other common subroutines.

Trios can naturally be extended to any multi-qubit operation of three or more qubits but this introduces the challenges of simultaneously routing many qubits and of designing decompositions that are efficient with whichever grouping the simultaneous router can achieve. It is not obvious how to route more than three qubits into a line or other desired shape.  As many NISQ benchmarks are not typically written with more complex structures and usually phrase them in terms of one-, two-, and three-qubit gates, this extension may only be desirable for larger-scale quantum computing.

\section{Evaluation}
\label{sec:evaluation}

\subsection{Toffoli Only Circuits}
We first evaluate the effect of our new compilation strategy by studying simple circuits containing only a single Toffoli gate. In these experiments, we place the three input qubits at random locations on the target hardware to emulate the potential locations of the qubits at some intermediate point in the execution of a more complex circuit. 

We study these circuits on a real IBM device, namely IBM Johannesburg, a 20-qubit device with limited connectivity in Figure \ref{fig:device-diagrams}a. We use the default Qiskit compiler which decomposes the Toffoli gates before doing shortest path routing compared to our proposed method where we do shortest path routing first and then decompose the Toffoli. We study the use of two different Toffoli implementations, a 6 CNOT decomposition with full qubit connectivity and an 8 CNOT decomposition with linear qubit connectivity.

In all four configurations, we compare the total compiled CNOT counts which correlates with the total success probability of a program. For execution on Johannesburg, we prepare the qubits in the states $\ket{110}$, perform the compiled Toffoli, then measure the three qubits of interest and compute the success rate as the probability of obtaining the correct answer (here the $\ket{111}$ state), where each experiment is performed with 8192 trials. 

\subsection{NISQ Benchmarks and Quantum Subroutines}

We also study Trios on real quantum benchmarks of moderate size using simulation only. The error rates of current devices are still too high to run benchmarks of these sizes but are expected to run on current devices as errors improve in the near future.  We choose error rates 20x better than Johannesburg rates as this make the estimated success probabilities within a reasonable range and is a realistic near-term estimate.  We discuss sensitivity to this choice later.

We study four implementations of the many-controlled-NOT (CnX) gate. This subroutine has many use cases from Grover's algorithm to various arithmetics. The implementations take advantage of differing numbers of ancilla and are chosen based on the number of available qubits on hardware. We study three adder implementations: Cuccaro, Takahashi, and QFT. The first two have many uses of the Toffoli gate while the latter has no such gates, for comparison. We study a small version of Grover's algorithm as well which makes use of the \verb|cnx_logancilla| subroutine. Finally, we compile two common NISQ benchmarks: QAOA for Max-Cut and Bernstein Vazirani (BV). We expect no gain on these benchmarks since they do not contain any Toffoli gates. A summary of our benchmarks is found in Table \ref{tab:benchmark_details} using implementations found in \cite{ourbenchmarks}.

\begin{table}[!b]
    \centering
    \begin{tabular}{cccc}
    \toprule
    Benchmark & Qubits & Toffolis & CNOTs* \\
    \midrule
    \verb|cnx_dirty| \cite{toff_any}               & 11 &  16 & 128 \\
    \verb|cnx_halfborrowed| \cite{gidney_toff}        & 19 &  32 & 256 \\
    \verb|cnx_logancilla| \cite{barenco}   & 19 &  17 & 136 \\
    \verb|cnx_inplace| \cite{gidney_toff}             &  4 &  54 & 490 \\
    \verb|cuccaro_adder| \cite{cuccaro}           & 20 &  18 & 190 \\
    \verb|takahashi_adder| \cite{takahashi}         & 20 &  18 & 188 \\
    \verb|incrementer_borrowedbit| \cite{gidney_toff} &  5 &  50 & 448 \\
    \verb|grovers|\cite{grover}                  &  9 &  84 & 672 \\
    \verb|qft_adder| \cite{qft}               & 16 &   0 &  92 \\
    \verb|bv| \cite{bv}                      & 20 &   0 &  19 \\
    \verb|qaoa_complete| \cite{qaoa}           & 10 &   0 &  90 \\
    \bottomrule
    \\
    \end{tabular}
    \caption{Details about our benchmarks both NISQ programs and other quantum subroutines. We consider circuits with and without Toffoli gates where we expect advantage only for circuits containing Toffoli gates. For BV we assume the all 1-bit string. The different CnX (many-controlled-NOT) gates use various numbers of ancilla. *The total number of CNOT gates is after decomposition with the 8-CNOT Toffoli but does not including any SWAPs for routing.}
    \label{tab:benchmark_details}
\end{table}

As noted previously, the connectivity of the underlying hardware has a significant impact on the number of required SWAPs. For example, on a completely connected set of qubits, no SWAPs are ever needed. In architectures with greater connectivity, we may opt for a more efficient Toffoli decomposition using 6 CNOTs. With simulation we study the effect of connectivity on the overall expected success rates and gate counts. We study four different connectivity models, all shown in Figure \ref{fig:device-diagrams}, each with 20 qubits, the topology of IBM's Johannesburg device containing four connected rings, a 2D mesh, a line, and a small clustered architecture representative of a QCCD ion trap.

We use error rates reported by IBM obtained via randomized benchmarking on a daily basis; for simulations we use error numbers obtained from Johannesburg obtained on 8/19/2020 with an average T1 time of $70.87\mu s$, T2 time of $72.72\mu s$, two qubit gate time of $0.559 \mu s$, a one qubit gate time of $0.07\mu s$, two qubit gate error of 0.0147, one qubit gate error of 0.0004. Source code for all experiments is available at \cite{sourcecode}.  Experiments using IBM are tested with version 0.14.0 through their Python API. When compiling with Qiskit for the single Toffoli experiments, we use the default settings for the \verb|transpile| function while specifying the Johannesburg backend. This means light optimization is performed: a stochastic routing policy is chosen, and some simple optimizations such as single qubit gate consolidation is performed. We fix the initial mapping to force routing to occur.
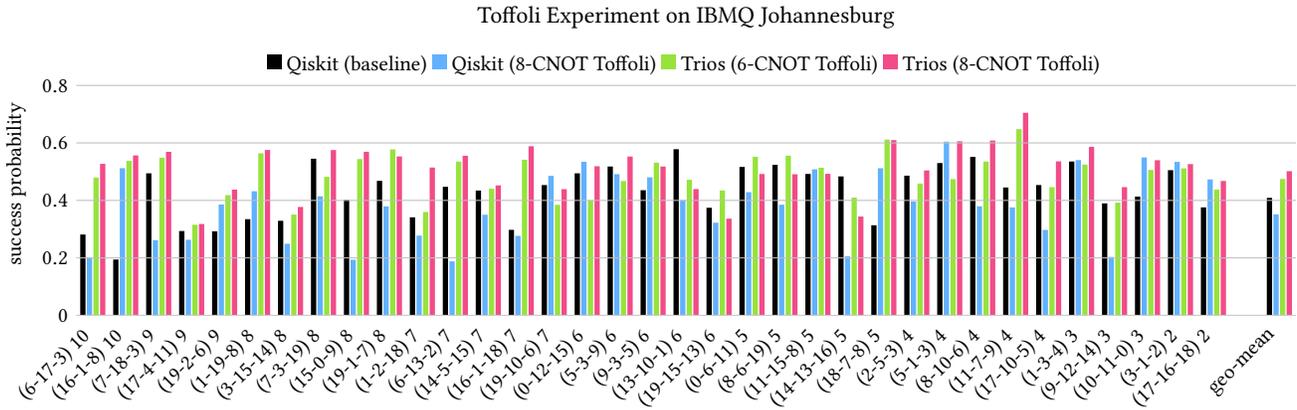
\begin{figure*}
    \centering
    \makebox[0.95\textwidth][r]{%
        \begin{tikzpicture}[baseline,scale=1,trim axis left,trim axis right]
\pgfplotsset{every tick label/.append style={font=\small}}
\pgfplotsset{every axis label/.append style={font=\small}}

    \begin{axis}[
        name=plot0,
        title={Toffoli Experiment on IBMQ Johannesburg},
        xlabel={},
        ylabel={success probability},
        symbolic x coords={(6-17-3) 10,(16-1-8) 10,(7-18-3) 9,(17-4-11) 9,(19-2-6) 9,(1-19-8) 8,(3-15-14) 8,(7-3-19) 8,(15-0-9) 8,(19-1-7) 8,(1-2-18) 7,(6-13-2) 7,(14-5-15) 7,(16-1-18) 7,(19-10-6) 7,(0-12-15) 6,(5-3-9) 6,(9-3-5) 6,(13-10-1) 6,(19-15-13) 6,(0-6-11) 5,(8-6-19) 5,(11-15-8) 5,(14-13-16) 5,(18-7-8) 5,(2-5-3) 4,(5-1-3) 4,(8-10-6) 4,(11-7-9) 4,(17-10-5) 4,(1-3-4) 3,(9-12-14) 3,(10-11-0) 3,(3-1-2) 2,(17-16-18) 2,gap,geo-mean},
        width={\linewidth},
        height={0.6*\columnwidth},
        ybar={0.5pt},
        bar width={2pt},
        enlargelimits=0.013888888888888888,
        ymin=0, ymax=0.9161254882812501,
        xtick=data,
        ,
        legend style={draw=none, fill=none, at={(0.5,1.03)},anchor=north,font=\small},
        legend columns=-1,
        legend image code/.code={\draw[#1, draw=none] (0em,-0.2em) rectangle (0.6em,0.4em);},
        axis line style={draw=black!20!white},
        axis on top,
        y axis line style={draw=none},
        axis x line*=bottom,
        tick style={draw=none},
        ,
        clip=false,
        enlarge y limits=0,
        ,
        x tick label style={rotate=45, anchor=east},
        grid=none,
        ymajorgrids=true,
        ,
        ,
        nodes near coords always on top/.style={
            every node near coord/.append style={
                anchor=south,
                rotate=0,
                font=\small,
                inner sep=0.2em,
            },
        },
        nodes near coords always on top,
    ]

        \addplot[
            style={
                color=transparent,
                draw=none,
                fill=black,
                ,
                mark=none,
                ,
                pattern color=black,,
            }]
        coordinates {
            ((6-17-3) 10, 0.2813720703125)
            ((16-1-8) 10, 0.1943359375)
            ((7-18-3) 9, 0.4942626953125)
            ((17-4-11) 9, 0.29345703125)
            ((19-2-6) 9, 0.2926025390625)
            ((1-19-8) 8, 0.334716796875)
            ((3-15-14) 8, 0.3291015625)
            ((7-3-19) 8, 0.5450439453125)
            ((15-0-9) 8, 0.4017333984375)
            ((19-1-7) 8, 0.468017578125)
            ((1-2-18) 7, 0.3411865234375)
            ((6-13-2) 7, 0.4473876953125)
            ((14-5-15) 7, 0.434326171875)
            ((16-1-18) 7, 0.297607421875)
            ((19-10-6) 7, 0.453369140625)
            ((0-12-15) 6, 0.4942626953125)
            ((5-3-9) 6, 0.5179443359375)
            ((9-3-5) 6, 0.435546875)
            ((13-10-1) 6, 0.578125)
            ((19-15-13) 6, 0.3743896484375)
            ((0-6-11) 5, 0.5164794921875)
            ((8-6-19) 5, 0.523681640625)
            ((11-15-8) 5, 0.4920654296875)
            ((14-13-16) 5, 0.4835205078125)
            ((18-7-8) 5, 0.3135986328125)
            ((2-5-3) 4, 0.486083984375)
            ((5-1-3) 4, 0.530029296875)
            ((8-10-6) 4, 0.5513916015625)
            ((11-7-9) 4, 0.4449462890625)
            ((17-10-5) 4, 0.4534912109375)
            ((1-3-4) 3, 0.535400390625)
            ((9-12-14) 3, 0.3896484375)
            ((10-11-0) 3, 0.4132080078125)
            ((3-1-2) 2, 0.5052490234375)
            ((17-16-18) 2, 0.375732421875)
            (gap, nan)
            (geo-mean, 0.4088172304790279)
        };
        \addlegendentry{Qiskit (baseline)~~~~};

        \addplot[
            style={
                color=transparent,
                draw=none,
                fill={rgb,190:red,72;green,132;blue,189},
                ,
                mark=none,
                ,
                pattern color={rgb,190:red,72;green,132;blue,189},,
            }]
        coordinates {
            ((6-17-3) 10, 0.1986083984375)
            ((16-1-8) 10, 0.5120849609375)
            ((7-18-3) 9, 0.26171875)
            ((17-4-11) 9, 0.26318359375)
            ((19-2-6) 9, 0.3856201171875)
            ((1-19-8) 8, 0.431884765625)
            ((3-15-14) 8, 0.249267578125)
            ((7-3-19) 8, 0.414306640625)
            ((15-0-9) 8, 0.193603515625)
            ((19-1-7) 8, 0.37939453125)
            ((1-2-18) 7, 0.2774658203125)
            ((6-13-2) 7, 0.18798828125)
            ((14-5-15) 7, 0.3502197265625)
            ((16-1-18) 7, 0.276611328125)
            ((19-10-6) 7, 0.4854736328125)
            ((0-12-15) 6, 0.5345458984375)
            ((5-3-9) 6, 0.4913330078125)
            ((9-3-5) 6, 0.4803466796875)
            ((13-10-1) 6, 0.4005126953125)
            ((19-15-13) 6, 0.322509765625)
            ((0-6-11) 5, 0.4285888671875)
            ((8-6-19) 5, 0.385009765625)
            ((11-15-8) 5, 0.5081787109375)
            ((14-13-16) 5, 0.2054443359375)
            ((18-7-8) 5, 0.511962890625)
            ((2-5-3) 4, 0.396240234375)
            ((5-1-3) 4, 0.6043701171875)
            ((8-10-6) 4, 0.3795166015625)
            ((11-7-9) 4, 0.3756103515625)
            ((17-10-5) 4, 0.29736328125)
            ((1-3-4) 3, 0.5404052734375)
            ((9-12-14) 3, 0.2032470703125)
            ((10-11-0) 3, 0.5494384765625)
            ((3-1-2) 2, 0.5340576171875)
            ((17-16-18) 2, 0.47314453125)
            (gap, nan)
            (geo-mean, 0.35173713914801175)
        };
        \addlegendentry{Qiskit (8-CNOT Toffoli)~~~~};

        \addplot[
            style={
                color=transparent,
                draw=none,
                fill={rgb,190:red,112;green,169;blue,45},
                ,
                mark=none,
                ,
                pattern color={rgb,190:red,112;green,169;blue,45},,
            }]
        coordinates {
            ((6-17-3) 10, 0.479248046875)
            ((16-1-8) 10, 0.537353515625)
            ((7-18-3) 9, 0.548583984375)
            ((17-4-11) 9, 0.315185546875)
            ((19-2-6) 9, 0.4183349609375)
            ((1-19-8) 8, 0.5638427734375)
            ((3-15-14) 8, 0.3505859375)
            ((7-3-19) 8, 0.4820556640625)
            ((15-0-9) 8, 0.5438232421875)
            ((19-1-7) 8, 0.5775146484375)
            ((1-2-18) 7, 0.3597412109375)
            ((6-13-2) 7, 0.5350341796875)
            ((14-5-15) 7, 0.440673828125)
            ((16-1-18) 7, 0.54150390625)
            ((19-10-6) 7, 0.3846435546875)
            ((0-12-15) 6, 0.3994140625)
            ((5-3-9) 6, 0.467529296875)
            ((9-3-5) 6, 0.5311279296875)
            ((13-10-1) 6, 0.4718017578125)
            ((19-15-13) 6, 0.4345703125)
            ((0-6-11) 5, 0.5518798828125)
            ((8-6-19) 5, 0.5556640625)
            ((11-15-8) 5, 0.5135498046875)
            ((14-13-16) 5, 0.409912109375)
            ((18-7-8) 5, 0.611572265625)
            ((2-5-3) 4, 0.4586181640625)
            ((5-1-3) 4, 0.4739990234375)
            ((8-10-6) 4, 0.5352783203125)
            ((11-7-9) 4, 0.64794921875)
            ((17-10-5) 4, 0.4459228515625)
            ((1-3-4) 3, 0.524658203125)
            ((9-12-14) 3, 0.392333984375)
            ((10-11-0) 3, 0.5059814453125)
            ((3-1-2) 2, 0.511474609375)
            ((17-16-18) 2, 0.4376220703125)
            (gap, nan)
            (geo-mean, 0.47450881060592054)
        };
        \addlegendentry{Trios (6-CNOT Toffoli)~~~~};

        \addplot[
            style={
                color=transparent,
                draw=none,
                fill={rgb,190:red,180;green,56;blue,101},
                ,
                mark=none,
                ,
                pattern color={rgb,190:red,180;green,56;blue,101},,
            }]
        coordinates {
            ((6-17-3) 10, 0.5272216796875)
            ((16-1-8) 10, 0.5565185546875)
            ((7-18-3) 9, 0.5689697265625)
            ((17-4-11) 9, 0.3179931640625)
            ((19-2-6) 9, 0.4375)
            ((1-19-8) 8, 0.5755615234375)
            ((3-15-14) 8, 0.377197265625)
            ((7-3-19) 8, 0.5753173828125)
            ((15-0-9) 8, 0.569091796875)
            ((19-1-7) 8, 0.552978515625)
            ((1-2-18) 7, 0.514404296875)
            ((6-13-2) 7, 0.5552978515625)
            ((14-5-15) 7, 0.4522705078125)
            ((16-1-18) 7, 0.58837890625)
            ((19-10-6) 7, 0.4390869140625)
            ((0-12-15) 6, 0.5189208984375)
            ((5-3-9) 6, 0.5526123046875)
            ((9-3-5) 6, 0.5179443359375)
            ((13-10-1) 6, 0.439453125)
            ((19-15-13) 6, 0.3365478515625)
            ((0-6-11) 5, 0.491943359375)
            ((8-6-19) 5, 0.490966796875)
            ((11-15-8) 5, 0.492431640625)
            ((14-13-16) 5, 0.3438720703125)
            ((18-7-8) 5, 0.6097412109375)
            ((2-5-3) 4, 0.5042724609375)
            ((5-1-3) 4, 0.605712890625)
            ((8-10-6) 4, 0.6082763671875)
            ((11-7-9) 4, 0.7047119140625)
            ((17-10-5) 4, 0.5357666015625)
            ((1-3-4) 3, 0.5863037109375)
            ((9-12-14) 3, 0.4461669921875)
            ((10-11-0) 3, 0.539794921875)
            ((3-1-2) 2, 0.5263671875)
            ((17-16-18) 2, 0.4676513671875)
            (gap, nan)
            (geo-mean, 0.5017832670104678)
        };
        \addlegendentry{Trios (8-CNOT Toffoli)};
    (

    \end{axis}

\end{tikzpicture}%
    }%
    \caption{Success probabilities of Toffoli gates between random triplets of qubits.  Higher is better.  The x labels specify the three qubits and total swap distance.  The geometric mean success rates for each compiler are 41\%, 35\%, 47\%, and 50\% respectively.  Trios (8-CNOT) improves average success rate by 23\% vs. the Qiskit baseline.}
    \label{fig:ibm-toffoli-success}
\end{figure*}

\begin{figure*}
    \centering
    \makebox[0.95\textwidth][r]{%
        \begin{tikzpicture}[baseline,scale=1,trim axis left,trim axis right]
\pgfplotsset{every tick label/.append style={font=\small}}
\pgfplotsset{every axis label/.append style={font=\small}}

    \begin{axis}[
        name=plot0,
        title={Toffoli Experiment on IBMQ Johannesburg},
        xlabel={},
        ylabel={CNOT gate count},
        symbolic x coords={(6-17-3) 10,(16-1-8) 10,(7-18-3) 9,(17-4-11) 9,(19-2-6) 9,(1-19-8) 8,(3-15-14) 8,(7-3-19) 8,(15-0-9) 8,(19-1-7) 8,(1-2-18) 7,(6-13-2) 7,(14-5-15) 7,(16-1-18) 7,(19-10-6) 7,(0-12-15) 6,(5-3-9) 6,(9-3-5) 6,(13-10-1) 6,(19-15-13) 6,(0-6-11) 5,(8-6-19) 5,(11-15-8) 5,(14-13-16) 5,(18-7-8) 5,(2-5-3) 4,(5-1-3) 4,(8-10-6) 4,(11-7-9) 4,(17-10-5) 4,(1-3-4) 3,(9-12-14) 3,(10-11-0) 3,(3-1-2) 2,(17-16-18) 2,gap,geo-mean},
        width={\linewidth},
        height={0.6*\columnwidth},
        ybar={0.5pt},
        bar width={2pt},
        enlargelimits=0.013888888888888888,
        ymin=0, ymax=70.2,
        xtick=data,
        ,
        legend style={draw=none, fill=none, at={(0.5,1.03)},anchor=north,font=\small},
        legend columns=-1,
        legend image code/.code={\draw[#1, draw=none] (0em,-0.2em) rectangle (0.6em,0.4em);},
        axis line style={draw=black!20!white},
        axis on top,
        y axis line style={draw=none},
        axis x line*=bottom,
        tick style={draw=none},
        ,
        clip=false,
        enlarge y limits=0,
        ,
        x tick label style={rotate=45, anchor=east},
        grid=none,
        ymajorgrids=true,
        ,
        ,
        nodes near coords always on top/.style={
            every node near coord/.append style={
                anchor=south,
                rotate=0,
                font=\small,
                inner sep=0.2em,
            },
        },
        nodes near coords always on top,
    ]

        \addplot[
            style={
                color=transparent,
                draw=none,
                fill=black,
                ,
                mark=none,
                ,
                pattern color=black,,
            }]
        coordinates {
            ((6-17-3) 10, 37)
            ((16-1-8) 10, 54)
            ((7-18-3) 9, 24)
            ((17-4-11) 9, 45)
            ((19-2-6) 9, 54)
            ((1-19-8) 8, 30)
            ((3-15-14) 8, 40)
            ((7-3-19) 8, 21)
            ((15-0-9) 8, 39)
            ((19-1-7) 8, 39)
            ((1-2-18) 7, 43)
            ((6-13-2) 7, 30)
            ((14-5-15) 7, 27)
            ((16-1-18) 7, 48)
            ((19-10-6) 7, 33)
            ((0-12-15) 6, 21)
            ((5-3-9) 6, 22)
            ((9-3-5) 6, 33)
            ((13-10-1) 6, 30)
            ((19-15-13) 6, 37)
            ((0-6-11) 5, 18)
            ((8-6-19) 5, 28)
            ((11-15-8) 5, 22)
            ((14-13-16) 5, 33)
            ((18-7-8) 5, 27)
            ((2-5-3) 4, 18)
            ((5-1-3) 4, 16)
            ((8-10-6) 4, 21)
            ((11-7-9) 4, 27)
            ((17-10-5) 4, 21)
            ((1-3-4) 3, 13)
            ((9-12-14) 3, 10)
            ((10-11-0) 3, 21)
            ((3-1-2) 2, 9)
            ((17-16-18) 2, 18)
            (gap, nan)
            (geo-mean, 28.965511018659207)
        };
        \addlegendentry{Qiskit (baseline)~~~~};

        \addplot[
            style={
                color=transparent,
                draw=none,
                fill={rgb,190:red,72;green,132;blue,189},
                ,
                mark=none,
                ,
                pattern color={rgb,190:red,72;green,132;blue,189},,
            }]
        coordinates {
            ((6-17-3) 10, 53)
            ((16-1-8) 10, 29)
            ((7-18-3) 9, 32)
            ((17-4-11) 9, 32)
            ((19-2-6) 9, 38)
            ((1-19-8) 8, 41)
            ((3-15-14) 8, 41)
            ((7-3-19) 8, 23)
            ((15-0-9) 8, 35)
            ((19-1-7) 8, 35)
            ((1-2-18) 7, 47)
            ((6-13-2) 7, 41)
            ((14-5-15) 7, 38)
            ((16-1-18) 7, 32)
            ((19-10-6) 7, 29)
            ((0-12-15) 6, 17)
            ((5-3-9) 6, 29)
            ((9-3-5) 6, 26)
            ((13-10-1) 6, 35)
            ((19-15-13) 6, 44)
            ((0-6-11) 5, 14)
            ((8-6-19) 5, 35)
            ((11-15-8) 5, 26)
            ((14-13-16) 5, 29)
            ((18-7-8) 5, 26)
            ((2-5-3) 4, 20)
            ((5-1-3) 4, 14)
            ((8-10-6) 4, 23)
            ((11-7-9) 4, 23)
            ((17-10-5) 4, 32)
            ((1-3-4) 3, 14)
            ((9-12-14) 3, 20)
            ((10-11-0) 3, 11)
            ((3-1-2) 2, 8)
            ((17-16-18) 2, 8)
            (gap, nan)
            (geo-mean, 27.571546147540772)
        };
        \addlegendentry{Qiskit (8-CNOT Toffoli)~~~~};

        \addplot[
            style={
                color=transparent,
                draw=none,
                fill={rgb,190:red,112;green,169;blue,45},
                ,
                mark=none,
                ,
                pattern color={rgb,190:red,112;green,169;blue,45},,
            }]
        coordinates {
            ((6-17-3) 10, 34)
            ((16-1-8) 10, 33)
            ((7-18-3) 9, 30)
            ((17-4-11) 9, 36)
            ((19-2-6) 9, 34)
            ((1-19-8) 8, 25)
            ((3-15-14) 8, 25)
            ((7-3-19) 8, 30)
            ((15-0-9) 8, 30)
            ((19-1-7) 8, 27)
            ((1-2-18) 7, 33)
            ((6-13-2) 7, 33)
            ((14-5-15) 7, 27)
            ((16-1-18) 7, 33)
            ((19-10-6) 7, 30)
            ((0-12-15) 6, 27)
            ((5-3-9) 6, 27)
            ((9-3-5) 6, 24)
            ((13-10-1) 6, 27)
            ((19-15-13) 6, 24)
            ((0-6-11) 5, 16)
            ((8-6-19) 5, 25)
            ((11-15-8) 5, 21)
            ((14-13-16) 5, 21)
            ((18-7-8) 5, 16)
            ((2-5-3) 4, 19)
            ((5-1-3) 4, 24)
            ((8-10-6) 4, 15)
            ((11-7-9) 4, 18)
            ((17-10-5) 4, 21)
            ((1-3-4) 3, 18)
            ((9-12-14) 3, 12)
            ((10-11-0) 3, 15)
            ((3-1-2) 2, 9)
            ((17-16-18) 2, 18)
            (gap, nan)
            (geo-mean, 23.39548267542428)
        };
        \addlegendentry{Trios (6-CNOT Toffoli)~~~~};

        \addplot[
            style={
                color=transparent,
                draw=none,
                fill={rgb,190:red,180;green,56;blue,101},
                ,
                mark=none,
                ,
                pattern color={rgb,190:red,180;green,56;blue,101},,
            }]
        coordinates {
            ((6-17-3) 10, 29)
            ((16-1-8) 10, 29)
            ((7-18-3) 9, 26)
            ((17-4-11) 9, 29)
            ((19-2-6) 9, 29)
            ((1-19-8) 8, 26)
            ((3-15-14) 8, 26)
            ((7-3-19) 8, 23)
            ((15-0-9) 8, 26)
            ((19-1-7) 8, 26)
            ((1-2-18) 7, 23)
            ((6-13-2) 7, 23)
            ((14-5-15) 7, 23)
            ((16-1-18) 7, 23)
            ((19-10-6) 7, 23)
            ((0-12-15) 6, 17)
            ((5-3-9) 6, 20)
            ((9-3-5) 6, 20)
            ((13-10-1) 6, 20)
            ((19-15-13) 6, 20)
            ((0-6-11) 5, 14)
            ((8-6-19) 5, 17)
            ((11-15-8) 5, 17)
            ((14-13-16) 5, 17)
            ((18-7-8) 5, 17)
            ((2-5-3) 4, 14)
            ((5-1-3) 4, 14)
            ((8-10-6) 4, 14)
            ((11-7-9) 4, 14)
            ((17-10-5) 4, 14)
            ((1-3-4) 3, 11)
            ((9-12-14) 3, 11)
            ((10-11-0) 3, 11)
            ((3-1-2) 2, 8)
            ((17-16-18) 2, 8)
            (gap, nan)
            (geo-mean, 18.824704952412723)
        };
        \addlegendentry{Trios (8-CNOT Toffoli)};
    (

    \end{axis}

\end{tikzpicture}%
    }%
    \caption{Total number of two-qubit (CNOT) gates required to execute a Toffoli gate between various distant qubits. Lower is better.  The x labels specify the three qubits and total swap distance.  The geometric mean gate counts for each compiler are 29, 28, 23, and 19 respectively.  Trios (8-CNOT) reduces average gate count by 35\%.}
    \label{fig:ibm-toffoli-gates}
\end{figure*}
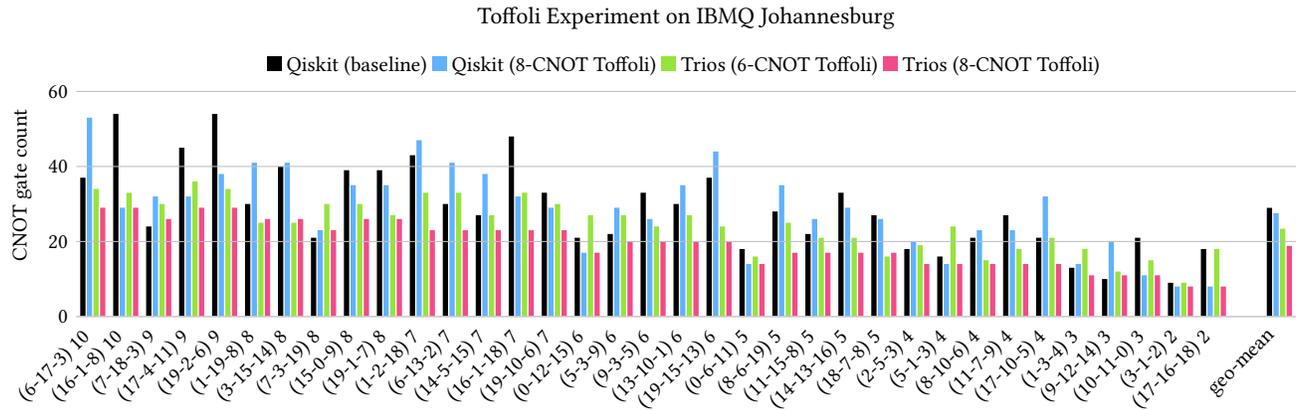

\begin{figure*}
    \centering
    \makebox[0.95\textwidth][r]{%
        \input{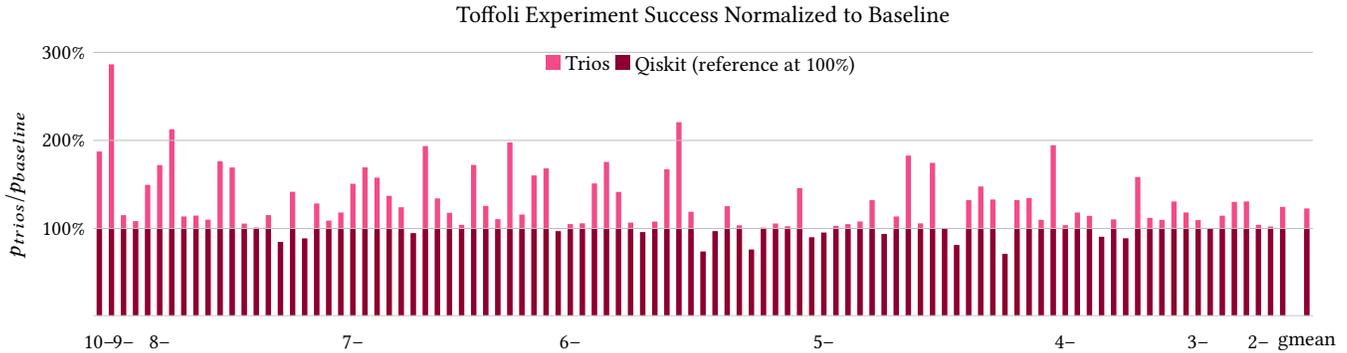}%
    }%
    \caption{Normalized success probabilities of Toffoli gates between triplets of qubits. Higher is better. Bars below 100\% indicate lower success rate for Trios.  The geometric mean increase in success rate is 23\%. The x labels indicate the qubit distance for a range of bars.}
    \label{fig:ibm-toffoli-success-norm}
\end{figure*}

\begin{figure*}
    \centering
    \makebox[0.95\textwidth][r]{%
        \input{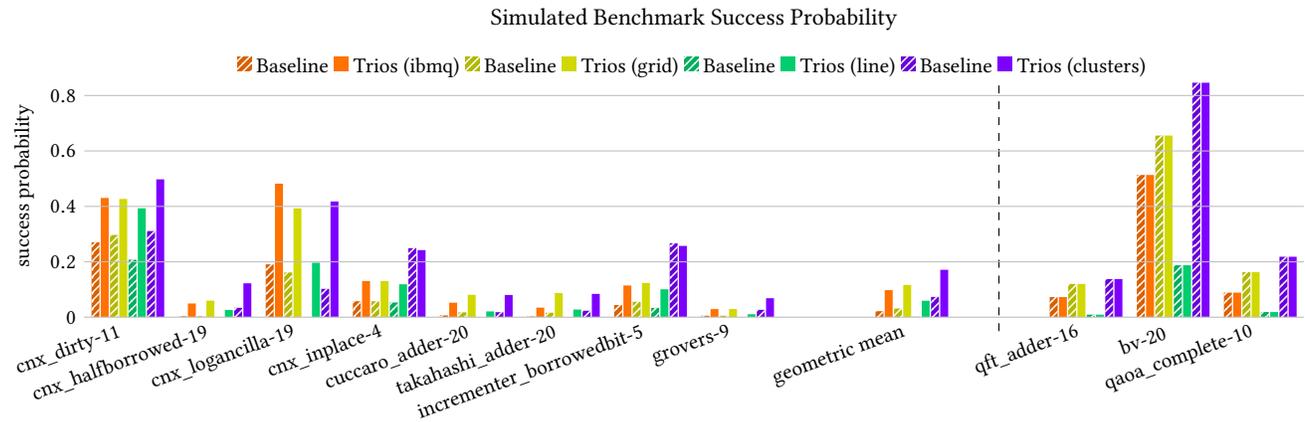}%
    }%
    \caption{Simulated upper-bounds on the program execution success probability on various hardware (using 20x lower idle and gate errors than Johannesburg).  Neighboring pairs of bars compare the baseline with Trios compiled for Johannesburg.  Higher is better when comparing pairs of bars with the same color.  The geometric mean success rates over the benchmarks that use Toffoli gate for each device type respectively are 2.2\%$\rightarrow$9.8\%, 3.2\%$\rightarrow$12\%, 0.19\%$\rightarrow$6.0\%, 7.3\%$\rightarrow$17\%.  The rightmost three benchmarks contain zero Toffoli gates so have no change vs. the baseline.}
    \label{fig:benchmark-success}
\end{figure*}

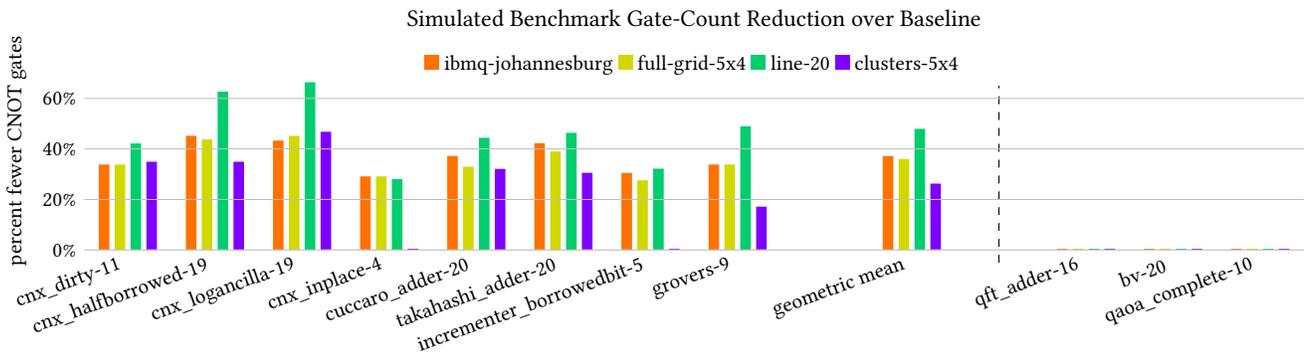
\begin{figure*}
    \centering
    \makebox[0.95\textwidth][r]{%
        \begin{tikzpicture}[baseline,scale=1,trim axis left,trim axis right]
\pgfplotsset{every tick label/.append style={font=\small}}
\pgfplotsset{every axis label/.append style={font=\small}}

    \begin{axis}[
        name=plot0,
        title={Simulated Benchmark Gate-Count Reduction over Baseline},
        xlabel={},
        ylabel={percent fewer CNOT gates},
        symbolic x coords={cnx\_dirty-11,cnx\_halfborrowed-19,cnx\_logancilla-19,cnx\_inplace-4,cuccaro\_adder-20,takahashi\_adder-20,incrementer\_borrowedbit-5,grovers-9, ,geometric mean,  ,qft\_adder-16,bv-20,qaoa\_complete-10},
        width={\linewidth},
        height={0.5*\columnwidth},
        ybar={2pt},
        bar width={4pt},
        enlargelimits=0.038461538461538464,
        ymin=0, ymax=79,
        xtick=data,
        ,
        legend style={draw=none, fill=none, at={(0.5,1.03)},anchor=north,font=\small},
        legend columns=-1,
        legend image code/.code={\draw[#1, draw=none] (0em,-0.2em) rectangle (0.6em,0.4em);},
        axis line style={draw=black!20!white},
        axis on top,
        y axis line style={draw=none},
        axis x line*=bottom,
        tick style={draw=none},
        yticklabel={\pgfmathparse{\tick*1}\pgfmathprintnumber{\pgfmathresult}\%},
        clip=false,
        enlarge y limits=0,
        ,
        x tick label style={rotate=20, anchor=east},
        grid=none,
        ymajorgrids=true,
        ,
        ,
        nodes near coords always on top/.style={
            every node near coord/.append style={
                anchor=south,
                rotate=0,
                font=\small,
                inner sep=0.2em,
            },
        },
        nodes near coords always on top,
    ]
        \addplot[
            style={
                color=transparent,
                draw=none,
                fill={rgb,255:red,255;green,113;blue,0},
                ,
                mark=none,
                ,
                pattern color={rgb,255:red,255;green,113;blue,0},,
            }]
        coordinates {
            (cnx\_dirty-11, 33.83333333333333)
            (cnx\_halfborrowed-19, 45.196969696969695)
            (cnx\_logancilla-19, 43.303030303030305)
            (cnx\_inplace-4, 29.147214854111407)
            (cuccaro\_adder-20, 37.234693877551024)
            (takahashi\_adder-20, 42.285714285714285)
            (incrementer\_borrowedbit-5, 30.525445292620866)
            (grovers-9, 33.83333333333333)
            ( , nan)
            (geometric mean, 37.17711233811081)
            (  , nan)
            (qft\_adder-16, 0.5)
            (bv-20, 0.5)
            (qaoa\_complete-10, 0.5)
        };
        \addlegendentry{ibmq-johannesburg~~~~};

        \addplot[
            style={
                color=transparent,
                draw=none,
                fill={rgb,255:red,209;green,216;blue,0},
                ,
                mark=none,
                ,
                pattern color={rgb,255:red,209;green,216;blue,0},,
            }]
        coordinates {
            (cnx\_dirty-11, 33.83333333333333)
            (cnx\_halfborrowed-19, 43.75396825396825)
            (cnx\_logancilla-19, 45.166666666666664)
            (cnx\_inplace-4, 29.147214854111407)
            (cuccaro\_adder-20, 32.97422680412371)
            (takahashi\_adder-20, 39.00267379679144)
            (incrementer\_borrowedbit-5, 27.600271002710024)
            (grovers-9, 33.83333333333333)
            ( , nan)
            (geometric mean, 35.949396509888224)
            (  , nan)
            (qft\_adder-16, 0.5)
            (bv-20, 0.5)
            (qaoa\_complete-10, 0.5)
        };
        \addlegendentry{full-grid-5x4~~~~};

        \addplot[
            style={
                color=transparent,
                draw=none,
                fill={rgb,255:red,0;green,205;blue,110},
                ,
                mark=none,
                ,
                pattern color={rgb,255:red,0;green,205;blue,110},,
            }]
        coordinates {
            (cnx\_dirty-11, 42.16666666666667)
            (cnx\_halfborrowed-19, 62.67948717948718)
            (cnx\_logancilla-19, 66.41880341880342)
            (cnx\_inplace-4, 28.05102040816326)
            (cuccaro\_adder-20, 44.40243902439025)
            (takahashi\_adder-20, 46.37378640776699)
            (incrementer\_borrowedbit-5, 32.2016317016317)
            (grovers-9, 48.984848484848484)
            ( , nan)
            (geometric mean, 47.945364137106395)
            (  , nan)
            (qft\_adder-16, 0.5)
            (bv-20, 0.5)
            (qaoa\_complete-10, 0.5)
        };
        \addlegendentry{line-20~~~~};

        \addplot[
            style={
                color=transparent,
                draw=none,
                fill={rgb,255:red,125;green,0;blue,255},
                ,
                mark=none,
                ,
                pattern color={rgb,255:red,125;green,0;blue,255},,
            }]
        coordinates {
            (cnx\_dirty-11, 34.94444444444444)
            (cnx\_halfborrowed-19, 34.94444444444444)
            (cnx\_logancilla-19, 46.7962962962963)
            (cnx\_inplace-4, 0.5)
            (cuccaro\_adder-20, 32.107629427792915)
            (takahashi\_adder-20, 30.561349693251532)
            (incrementer\_borrowedbit-5, 0.5)
            (grovers-9, 17.166666666666664)
            ( , nan)
            (geometric mean, 26.288559342717726)
            (  , nan)
            (qft\_adder-16, 0.5)
            (bv-20, 0.5)
            (qaoa\_complete-10, 0.5)
        };
        \addlegendentry{clusters-5x4};
    \addplot[draw=black,dashed,smooth]
    coordinates {(  ,-5) (  ,65)};%
    \end{axis}

\end{tikzpicture}%
    }%
    \caption{A comparison between the baseline and Trios for various hardware.  Above 0\% indicates benefit.  All two-qubit gates (for communication and computation) are counted.  The geometric mean reductions in gate counts are 37\%, 36\%, 48\%, and 26\% respectively.  The rightmost three benchmarks contain zero Toffoli gates so have no change vs. the baseline.}
    \label{fig:benchmark-gate-ratio}
\end{figure*}

\begin{figure*}
    \centering
    \makebox[0.95\textwidth][r]{%
        \begin{tikzpicture}[baseline,scale=1,trim axis left,trim axis right]
\pgfplotsset{every tick label/.append style={font=\small}}
\pgfplotsset{every axis label/.append style={font=\small}}

    \begin{semilogyaxis}[
        name=plot0,
        title={Simulated Benchmark Success Normalized to Baseline},
        xlabel={},
        ylabel={},
        symbolic x coords={cnx\_dirty-11,cnx\_halfborrowed-19,cnx\_logancilla-19,cnx\_inplace-4,cuccaro\_adder-20,takahashi\_adder-20,incrementer\_borrowedbit-5,grovers-9, ,geometric mean,  ,qft\_adder-16,bv-20,qaoa\_complete-10},
        width={\linewidth},
        height={0.5*\columnwidth},
        ybar={1.5pt},
        bar width={4pt},
        enlargelimits=0.038461538461538464,
        ymin=1, ymax=2999,
        xtick=data,
        ,
        legend style={draw=none, fill=none, at={(0.5,1.03)},anchor=north,font=\small},
        legend columns=-1,
        legend image code/.code={\draw[#1, draw=none] (0em,-0.2em) rectangle (0.6em,0.4em);},
        axis line style={draw=black!20!white},
        axis on top,
        y axis line style={draw=none},
        axis x line*=bottom,
        tick style={draw=none},
        ,
        clip=false,
        enlarge y limits=0,
        ,
        x tick label style={rotate=20, anchor=east},
        grid=none,
        ymajorgrids=true,
        yminorgrids=true, ylabel=$p_{trios}/p_{baseline}$,
        ,
        nodes near coords always on top/.style={
            every node near coord/.append style={
                anchor=south,
                rotate=0,
                font=\small,
                inner sep=0.2em,
            },
        },
        nodes near coords always on top,
    ]
        \addplot[
            style={
                color=transparent,
                draw=none,
                fill={rgb,255:red,255;green,113;blue,0},
                ,
                mark=none,
                ,
                pattern color={rgb,255:red,255;green,113;blue,0},,
            }]
        coordinates {
            (cnx\_dirty-11, 1.5896648545735392)
            (cnx\_halfborrowed-19, 13.407651193733729)
            (cnx\_logancilla-19, 2.524547907696733)
            (cnx\_inplace-4, 2.263988613645191)
            (cuccaro\_adder-20, 7.6082931085593835)
            (takahashi\_adder-20, 11.80367295443326)
            (incrementer\_borrowedbit-5, 2.597857631656044)
            (grovers-9, 5.326082341226377)
            ( , nan)
            (geometric mean, 4.441252548195526)
            (  , nan)
            (qft\_adder-16, 1.0)
            (bv-20, 1.0)
            (qaoa\_complete-10, 1.0)
        };
        \addlegendentry{ibmq-johannesburg~~~~};

        \addplot[
            style={
                color=transparent,
                draw=none,
                fill={rgb,255:red,209;green,216;blue,0},
                ,
                mark=none,
                ,
                pattern color={rgb,255:red,209;green,216;blue,0},,
            }]
        coordinates {
            (cnx\_dirty-11, 1.4377463977511804)
            (cnx\_halfborrowed-19, 15.311806031497524)
            (cnx\_logancilla-19, 2.4307232746263026)
            (cnx\_inplace-4, 2.263988613645191)
            (cuccaro\_adder-20, 4.521922465526685)
            (takahashi\_adder-20, 5.356175676310816)
            (incrementer\_borrowedbit-5, 2.2185736795499724)
            (grovers-9, 5.326082341226377)
            ( , nan)
            (geometric mean, 3.6940031823077475)
            (  , nan)
            (qft\_adder-16, 1.0)
            (bv-20, 1.0)
            (qaoa\_complete-10, 1.0)
        };
        \addlegendentry{full-grid-5x4~~~~};

        \addplot[
            style={
                color=transparent,
                draw=none,
                fill={rgb,255:red,0;green,205;blue,110},
                ,
                mark=none,
                ,
                pattern color={rgb,255:red,0;green,205;blue,110},,
            }]
        coordinates {
            (cnx\_dirty-11, 1.8933437282657255)
            (cnx\_halfborrowed-19, 4258.64023330463)
            (cnx\_logancilla-19, 201.07417925378348)
            (cnx\_inplace-4, 2.227549686978761)
            (cuccaro\_adder-20, 23.224060039305026)
            (takahashi\_adder-20, 39.4009760141775)
            (incrementer\_borrowedbit-5, 2.9791791316994827)
            (grovers-9, 90.98391900034488)
            ( , nan)
            (geometric mean, 31.19060603069423)
            (  , nan)
            (qft\_adder-16, 1.0)
            (bv-20, 1.0)
            (qaoa\_complete-10, 1.0)
        };
        \addlegendentry{line-20~~~~};

        \addplot[
            style={
                color=transparent,
                draw=none,
                fill={rgb,255:red,125;green,0;blue,255},
                ,
                mark=none,
                ,
                pattern color={rgb,255:red,125;green,0;blue,255},,
            }]
        coordinates {
            (cnx\_dirty-11, 1.599641130404093)
            (cnx\_halfborrowed-19, 3.5894014754239083)
            (cnx\_logancilla-19, 4.074161573419985)
            (cnx\_inplace-4, 0.9725673948512388)
            (cuccaro\_adder-20, 4.379170294963949)
            (takahashi\_adder-20, 3.565238909329951)
            (incrementer\_borrowedbit-5, 0.9643362774598649)
            (grovers-9, 2.554455092114703)
            ( , nan)
            (geometric mean, 2.3321217818921745)
            (  , nan)
            (qft\_adder-16, 1.0)
            (bv-20, 1.0)
            (qaoa\_complete-10, 1.0)
        };
        \addlegendentry{clusters-5x4};
    \addplot[draw=black,dashed,smooth]
    coordinates {(  ,0.6) (  ,1000)};%
    \end{semilogyaxis}

\end{tikzpicture}%
    }%
    \caption{Normalized Figure \ref{fig:benchmark-success} to show our consistent increase in program success with Trios.    Above $10^0$ indicates benefit.  Some improvement factors are huge due to near-zero baseline success rates.  The geometric mean increases in success rate are 4.4x, 3.7x, 31x, and 2.3x respectively.  The rightmost three benchmarks contain zero Toffoli gates so have no change vs. the baseline.}
    \label{fig:benchmark-success-norm}
\end{figure*}
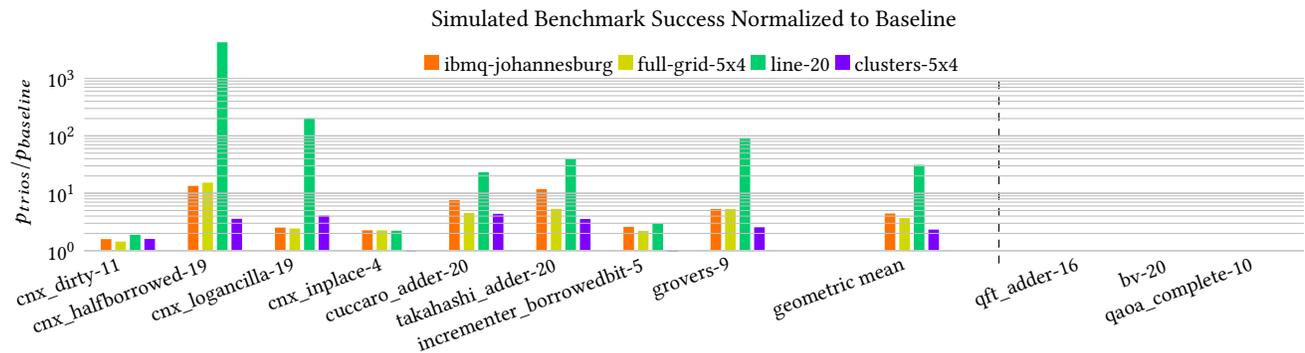

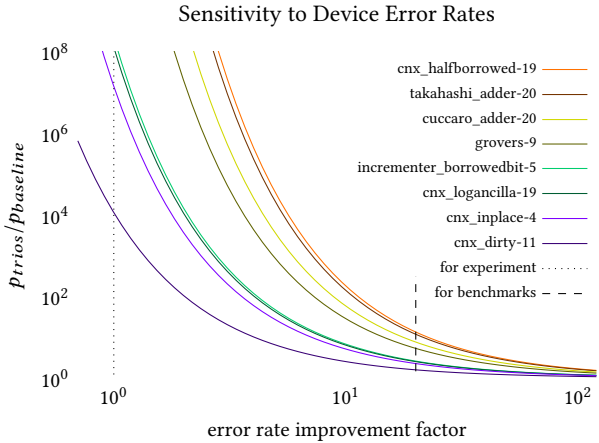
\begin{figure}
    \centering
    \makebox[0.95\columnwidth][r]{%
        \begin{tikzpicture}[baseline,scale=1,trim axis left,trim axis right]
\pgfplotsset{every tick label/.append style={font=\small}}
\pgfplotsset{every axis label/.append style={font=\small}}

    \begin{loglogaxis}[
        name=plot9,
        title={Sensitivity to Device Error Rates},
        xlabel={error rate improvement factor},
        ylabel={},
        width={\columnwidth},
        height={0.7*\columnwidth},
        xmin=0.7, xmax=120, ymin=0.99, ymax=102329299.22807537,
        ,
        legend style={
            draw=none,
            at={(1,1)},
            anchor=north east,
            font=\small},
        ,
        clip=false,
        axis line style={draw=none},
        tick style={draw=none},,
        ylabel=$p_{trios}/p_{baseline}$,
    legend style={
        cells={anchor=east},
        legend plot pos=right,
        draw=none,
        at={(1,1)},
        anchor=north east,
        fill=none,
        font=\small},
    clip=true,
    ]
        \addplot[color={rgb,255:red,255;green,113;blue,0}] table[x=error-rate-improvement-factor 0, y=scriptsizecnxhalfborrowed-19  0, col sep=comma]
            {data/sensitivity-to-device-error-rates.csv}
        ;%
        \addlegendentry{\scriptsize{cnx\_halfborrowed-19}};

        \addplot[color={rgb,555:red,255;green,113;blue,0}] table[x=error-rate-improvement-factor 1, y=scriptsizetakahashiadder-20  1, col sep=comma]
            {data/sensitivity-to-device-error-rates.csv}
        ;%
        \addlegendentry{\scriptsize{takahashi\_adder-20}};

        \addplot[color={rgb,255:red,209;green,216;blue,0}] table[x=error-rate-improvement-factor 2, y=scriptsizecuccaroadder-20  2, col sep=comma]
            {data/sensitivity-to-device-error-rates.csv}
        ;%
        \addlegendentry{\scriptsize{cuccaro\_adder-20}};

        \addplot[color={rgb,555:red,209;green,216;blue,0}] table[x=error-rate-improvement-factor 3, y=scriptsizegrovers-9  3, col sep=comma]
            {data/sensitivity-to-device-error-rates.csv}
        ;%
        \addlegendentry{\scriptsize{grovers-9}};

        \addplot[color={rgb,255:red,0;green,205;blue,110}] table[x=error-rate-improvement-factor 4, y=scriptsizeincrementerborrowedbit-5  4, col sep=comma]
            {data/sensitivity-to-device-error-rates.csv}
        ;%
        \addlegendentry{\scriptsize{incrementer\_borrowedbit-5}};

        \addplot[color={rgb,555:red,0;green,205;blue,110}] table[x=error-rate-improvement-factor 5, y=scriptsizecnxlogancilla-19  5, col sep=comma]
            {data/sensitivity-to-device-error-rates.csv}
        ;%
        \addlegendentry{\scriptsize{cnx\_logancilla-19}};

        \addplot[color={rgb,255:red,125;green,0;blue,255}] table[x=error-rate-improvement-factor 6, y=scriptsizecnxinplace-4  6, col sep=comma]
            {data/sensitivity-to-device-error-rates.csv}
        ;%
        \addlegendentry{\scriptsize{cnx\_inplace-4}};

        \addplot[color={rgb,555:red,125;green,0;blue,255}] table[x=error-rate-improvement-factor 7, y=scriptsizecnxdirty-11  7, col sep=comma]
            {data/sensitivity-to-device-error-rates.csv}
        ;%
        \addlegendentry{\scriptsize{cnx\_dirty-11}};

        \addplot[color=black, dotted] table[x=error-rate-improvement-factor 8, y=scriptsizefor-experiment  8, col sep=comma]
            {data/sensitivity-to-device-error-rates.csv}
        ;%
        \addlegendentry{\scriptsize{for experiment}};

        \addplot[color=black, dashed] table[x=error-rate-improvement-factor 9, y=scriptsizefor-benchmarks  9, col sep=comma]
            {data/sensitivity-to-device-error-rates.csv}
        ;%
        \addlegendentry{\scriptsize{for benchmarks}};

    \end{loglogaxis}

\end{tikzpicture}%
    }%
    \caption{Factor of improvement in success rate in Trios over baseline for scaling gate error rates. The dotted line indicates current error rates on IBM Johannesburg and the dashed line (20x improvement) indicates values of the near future used in simulation. In our approximation of success rate factors of improvement in gate error rates lead to an exponential fall off in success ratios, as expected. In the very near term, we expect Trios to drastically improve the execution of quantum programs.}
    \label{fig:benchmark-error-sensitivity}
\end{figure}

\section{Results and Discussion}
\label{sec:results}
\subsection{Trios Reduces Total Number of Gates}
In both sets of experiments, the total number of gates required to make the input programs executable is much less than when using the default Qiskit compiler. When compiling our simple programs consisting of a single Toffoli gate with qubits mapped in random locations, we reduce the average number of gates by 35\% geomean. 

In Figure \ref{fig:ibm-toffoli-gates} we show 35 different triplets of hardware qubits for each of the four strategies. For each triplet, we note the total distance between the qubits on the hardware, given by the shortest path distance in the underlying topology. Even when the distance is relatively small, Trios outperforms reducing overall gate count and as the distance increases, the margin tends to increase. In the small distance cases, this can be attributed to Trios choosing the better Toffoli decomposition for a linearly connected topology. This is significant for two reasons. First, the fewer the gates, the less likely an error occurs due to qubit manipulation. Second, fewer gates, especially long sequential chains of SWAPs, often means lower circuit depth, meaning fewer chances for decoherence errors. Together this translates into faster and more successful programs. 
 
This advantage extends to our NISQ benchmarks which contain various numbers of Toffoli gates. In Figure \ref{fig:benchmark-gate-ratio} we note substantial reductions in total gates across all benchmarks containing Toffoli gates across all underlying topologies. The only exception is the two smallest benchmarks (on 4 and 5 qubits) for the clustered topology because they could be compiled with zero SWAPs.

An extreme of the clustered topology is a single cluster with all-to-all connected qubits.  On this device, Orchestrated Trios would have no benefit as operations can be performed between any pair of qubits so no SWAPs are needed and routing is trivial.  However, as quantum technologies scale to more than a few qubits, fully-connected architectures hits physical limitations and must be re-engineered.  As trapped ion qubit chains get longer, for example, gate operations become slower and lower fidelity.  \cite{trapsize} showed that the optimal trap size is 15-25 ions interconnected similar to our cluster model with cluster sizes of 15-25 where Trios does benefit.

On average, for Toffoli-containing programs we reduce gate count 37\%, 36\%, 48\%, 26\% for Johannesburg, Grid, Line, and Cluster topologies respectively with the maximum gain obtained for linear devices. 

\subsection{Trios Improves Overall Success Rate}
In general, we expect programs with fewer total two-qubit gates, to succeed with higher probability. In devices with limited connectivity, the addition of routing operations like SWAPs, usually decomposed to 3 CNOTs, can severely reduce the chance an input program can succeed. While success rate is inversely correlated with number of gates, gate error is not the only reason a program can fail and reducing gate counts does not \textit{guarantee} improved success rates.

In Figure \ref{fig:ibm-toffoli-success} we show the success rates of our Toffoli-only experiments when the two controls are initialized to $\ket{1}$ and the target is initialized to $\ket{0}$ so we measure the probability of obtaining $\ket{111}$. These results are obtained from Johannesburg on 8/19/2020. The x-axes of both Figures \ref{fig:ibm-toffoli-success} and \ref{fig:ibm-toffoli-gates} line up to compare gate counts and resulting success rate. In general, experimentally, fewer gates results in substantial improvements to success rates. For example, a Toffoli on (6-17-3) compiled with Trios improves success rate from around 30\% to over 50\%. On average, we improve success rates by 23 \% geomean with max of 286\%. In Figure \ref{fig:ibm-toffoli-success-norm}, we show improvements compiled with Trios normalized to baseline for 99 different triplets of varying total distance on Johannesburg. 

Trios on average improves the probability of success for these circuits. However, there are a small number of cases where Trios performs worse despite having a smaller number of total gates. This can be attributed to several different factors. For example, the chosen edges for SWAP paths may be more noisy, or on pairs of edges with greater crosstalk, or the final qubits which are measured have worse readout error. Regardless, reducing the overall gate count of a program is an important contributing factor to improving expected success rate.

For our simulated NISQ benchmarks, we see even larger gains. The reduced gate counts in Figure \ref{fig:benchmark-gate-ratio} translate to major improvements in simulated success rate in Figure \ref{fig:benchmark-success} (normalized success rates in Figure \ref{fig:benchmark-success-norm}). For example, in \verb|cnx_logancilla-19|, Trios more than doubles the expected success rates when compiled to each of the architectures. In many cases, the expected success rate of programs compiled with Qiskit is effectively zero while Trios has a realistic chance of obtaining the correct answer. As expected, on programs containing no Toffoli gates, Trios has no effect on success showing that it introduces no excessive overhead. This suggests Trios can easily be added to other quantum compilation toolflows.

\subsection{Trios Routes Complex Interactions Better}
Trios improves gate counts, and consequently improves success rates, by routing more efficiently and choosing more appropriate Toffoli decompositions based on the underlying architecture's connectivity. Current compilers, like Qiskit, perform routing on fully decomposed and unrolled programs, and while this must eventually be done, it leads to less efficient routing policies and relies on assumptions that a theoretically good decomposition (fewer CNOTs) is the best decomposition for the hardware. Trios eliminates this by choosing a context-dependent Toffoli decomposition and routing multiqubit gates as single units.

Trios greatly improves effectiveness compared to a \textit{heuristic-based} compiler by applying similar heuristics to the higher abstraction level Toffoli gates.  An optimal routing of the decomposed circuit would be better except it cannot select the best architecture location-specific decomposition.  This makes a huge difference specifically with Toffolis on any square-grid-based device.  One might choose to improve the solution found by an optimal compiler by always decomposing Toffolis to the 8-CNOT version before optimally routing, but this will still limit the solution.  There are multiple possible qubit orders for the decomposition and the best can only be selected after the routing pass.

\balance
\subsection{Simulation Sensitivity to Error Rates}
For our simulations we use an error model (20x better than current errors on Johannesburg) which is forward looking. As errors improve, we expect Trios to have a reduced impact on program success rates since gate errors will contribute less and less to program failure though Trios will never perform worse than the baseline. In Figure \ref{fig:benchmark-error-sensitivity} we study the sensitivity of simulation results to two qubit error rates beginning with current IBM error rates. For poor error rates, the benefit of Trios is extremely large, owed to the fact that programs compiled with the baseline have probabilities of success very close to 0. In our simplified simulation framework, as error rates improve we expect an exponential drop off in improvement with the most advantage obtained with current error rates.

\vspace{3.25pt}
\section{Conclusion}
\label{sec:conclusion}
We present a new quantum compilation structure, Trios, with a split decomposition pass to greatly reduce compiled communication cost and enable architecture-tuned decompositions. We specifically target the three-qubit Toffoli operation to capture program structure enabling more optimal compiled circuits.  Because current quantum computers are especially error prone, they require high levels of optimization to reduce gate counts and maximize the probability the compiled program will succeed.

Orchestrated Trios both greatly improves the effectiveness of qubit routing given newly exposed program structure and improves decompositions with connectivity-awareness.  These both greatly benefit the program success rate, a critical metric for today's error-prone and resource-constrained quantum computers.  We hope this inspires more hierarchically designed NISQ algorithms now that we have shown breaking the abstractions of discrete compilation passes can help bridge the gap between these noisy quantum hardware and practical applications.

\vspace{3.25pt}
\begin{acks}
This work is funded in part by EPiQC, an NSF Expedition in Computing, under grants CCF-1730449; in part by STAQ under grant NSF Phy-1818914; in part by DOE grants DE-SC0020289 and DE-SC0020331; and in part by NSF OMA-2016136 and the Q-NEXT DOE NQI Center.

This research used resources of the Oak Ridge Leadership Computing Facility, which is a DOE Office of Science User Facility supported under Contract DE-AC05-00OR22725.

Disclosure: F. Chong is also Chief Scientist at Super.tech and an advisor to Quantum Circuits, Inc.
\end{acks}

\newpage

\bibliographystyle{ACM-Reference-Format}
\bibliography{refs}

\end{document}